\documentclass[iop,numberedappendix,twocolappendix,twocolumn]{emulateapj} 

\usepackage{latexsym}    
\usepackage{amsfonts}    
\usepackage{amstext}     
\usepackage{amsmath,amssymb}
\usepackage{mathtools}
\usepackage{bm}
\usepackage{MnSymbol} 
\usepackage{mathrsfs}
\usepackage{xcolor}
\usepackage{hyperref}
\hypersetup{
  colorlinks   = true, 
  urlcolor     = blue, 
  linkcolor    = blue, 
  citecolor   = blue 
}

\providecommand{\Psie}{\mathbf{\Psi}^E}
\providecommand{\Psia}{\mathbf{\Psi}^A}
\providecommand{\Psil}{\mathbf{\Psi}^L}

\begin{document}
 
\title{\bf Formal solutions for polarized radiative transfer\\ III. Stiffness and instability}
\author{Gioele Janett\altaffilmark{1,2}, Alberto Paganini\altaffilmark{3}} 
\email{gioele.janett@irsol.ch} 

\affil{$^1$ Istituto Ricerche Solari Locarno (IRSOL), 6605 Locarno-Monti, Switzerland\\
$^2$ Seminar for Applied Mathematics (SAM) ETHZ, 8093 Zurich, Switzerland\\
$^3$ University of Oxford, Mathematical Institute, OX2 6GG Oxford, United Kingdom}

\begin{abstract}
Efficient numerical approximation of the polarized radiative transfer equation is challenging
because this system of ordinary differential equations exhibits stiff behavior,
which potentially results in numerical instability.
This negatively impacts the accuracy of formal solvers, and small step-sizes are often necessary to retrieve physical solutions.
This work presents stability analyses
of formal solvers for the radiative transfer equation of polarized light, identifies instability issues, and suggests practical remedies.
In particular, the assumptions and the limitations of the stability analysis of Runge-Kutta methods play a crucial role.
On this basis, a suitable and pragmatic formal solver is outlined and tested.
An insightful comparison to the scalar radiative transfer equation is also presented.
\end{abstract}
\section{Introduction}\label{sec:intro}
The transfer of partially polarized light is described by the following linear system of first-order coupled inhomogeneous ODEs
 %
\begin{equation}
  \frac{\rm d}{{\rm d} s}\mathbf I(s) 
  = -\mathbf K(s)\mathbf I(s) + \boldsymbol{\epsilon}(s)\,,
\label{eq:RTE}
\end{equation}
where $s$ is the spatial coordinate measured along the ray under consideration, $\mathbf{I}$ is the Stokes vector, $\mathbf{K}$ is the propagation matrix, and $\boldsymbol{\epsilon}$ is the emission vector. For notational simplicity, the frequency dependence of these quantities is not explicitly indicated. 

It is common practice to solve Equation~\eqref{eq:RTE} by means of numerical methods,
because its analytical solution is known for a few simple atmospheric models (which determine $\mathbf K$ and $\boldsymbol{\epsilon}$) only.
 %
 %
 %
However, Equation~\eqref{eq:RTE} exhibits stiff behavior, i.e., formal solvers may face instability issues.
 %
For instance, \citet{murphy1990} observed instability problems using the DELO-parabolic method.  
Thereafter, \citet{bellot_rubio+al1998} encountered instability when using the cubic Hermitian method for the spectral synthesis of strong lines. \citet{delacruz_rodriguez+piskunov2013} underlined the importance of preserving stability when DELO methods are extended to high-order schemes in terms of quadratic and cubic B\'ezier interpolations. \citet{stepan+trujillo_bueno2013} use B\'ezier interpolants to control abrupt changes in the atmospheric quantities,
which potentially lead to instabilities.  
\citet{steiner2016} proposed a different approach to deal with strong gradients, using piecewise continuous reconstructions and slope limiters.
Finally, \citet{janett2017a,janett2017b}
provide a characterization of formal solvers in terms of their stability region paying particular attention
to the eigenvalues of the propagation matrix.


The concept and the relevance of stability are ubiquitous in numerical analysis, and numerical methods for ODEs
are not an exception  
\citep[e.g.,][]{dahlquist1963,deuflhard2002}.  
In particular, stability is a necessary condition for convergence. Indeed,
to ensure that a numerical solution of an ODE converges,
it is first necessary to show that the numerical scheme employed is consistent, that is,
that the local error introduced in one step decays superlinearly with respect to the step-size
$\Delta t$. Unfortunately, this consistency condition is not sufficient to ensure convergence
because the cumulative sum of local errors may grow exponentially. However, this exponential growth
cannot happen if the numerical method is stable. In light of this, \citet{hackbusch2014} concludes that
``whether consistency implies convergence depends on stability''.
Stability analysis is
employed to provide additional requirements to numerical methods (e.g., a limited step-size).
However, these particular stability requirements are problem-dependent and often difficult to be determined.

This paper aims to give a deeper analysis on stability conditions, when facing the numerical integration of Equation~\eqref{eq:RTE}. Section~\ref{sec:propmat} focuses on the propagation matrix and on its eigenvalues. Section~\ref{sec:stability} presents the stability analysis of 
Runge-Kutta methods.  
Particular attention is paid to the assumptions and the limitations of this analysis, emphasizing their relevance in the formal solution for polarized light.
Section~\ref{sec:conversion} analyzes the effect of the conversion to optical depth on numerical stability, while Section~\ref{sec:numericalconversion} exposes the numerical approximation of this conversion. Section~\ref{sec:bestsolver} describes the structure of a pragmatic numerical method for the numerical integration of Equation~\eqref{eq:RTE}.
Section~\ref{sec:sec7} presents complementary
considerations on this topic.
Finally, Section~\ref{sec:sec8} provides remarks and conclusions.
%
\section{The propagation matrix}\label{sec:propmat}
The propagation matrix $\mathbf K$ in Equation~\eqref{eq:RTE} can be written in the form \citep{landi_deglinnocenti+landolfi2004}
\begin{equation}
  \mathbf K = \begin{pmatrix}
      \eta_I &  \eta_Q &  \eta_U & \eta_V  \\
      \eta_Q &  \eta_I &  \rho_V & -\rho_U \\
      \eta_U & -\rho_V &  \eta_I & \rho_Q  \\
      \eta_V &  \rho_U & -\rho_Q & \eta_I 
              \end{pmatrix}\,,
\label{propagation_matrix}
              \end{equation}
where the seven independent coefficients are, in general, functions of the frequency,
propagation direction,  
and of a series of physical parameters describing the atmosphere. The matrix $\mathbf K$ can be decomposed into three different contributions, namely,
{\small
\begin{equation*}
  \begin{pmatrix}
      \eta_I & 0      & 0      & 0      \\
      0      & \eta_I & 0      & 0      \\
      0      & 0      & \eta_I & 0      \\
      0      & 0      & 0      & \eta_I 
  \end{pmatrix}+
  \begin{pmatrix}
      0      &  \eta_Q &  \eta_U & \eta_V  \\
      \eta_Q &  0      &  0      & 0       \\
      \eta_U &  0      &  0      & 0       \\
      \eta_V &  0      &  0      & 0      
  \end{pmatrix}+
  \begin{pmatrix}
      0      &  0      &  0      & 0       \\
      0      &  0      &  \rho_V & -\rho_U \\
      0      & -\rho_V &  0      & \rho_Q  \\
      0      &  \rho_U & -\rho_Q & 0      
  \end{pmatrix}\,.
\end{equation*}}\noindent
The first matrix is called the absorption matrix, it is diagonal, and it is responsible for the usual exponential decay of the whole Stokes vector.
The second matrix is called the dichroism matrix, it is symmetric, and it is responsible for dichroism effects, i.e.,
the property of absorbing light to different extents depending on the polarization states.
The third matrix is called the dispersion matrix, it is skew-symmetric, and it describes the coupling of the Stokes components due to anomalous dispersion effects. 

The propagation matrix coefficients consist, in general, of two different kinds of contributions: continuum processes (due to bound-free and free-free transitions) and
spectral lines (due to bound-bound transitions). In solar context, continuum processes do
not introduce dichroism or anomalous dispersion effects.
This section describes the propagation matrix coefficients for an isolated spectral line originating from the 
atomic transition between two levels with total angular momentum $J_u$ (upper level) and $J_\ell$ (lower level), respectively.
Each $J$-level is composed of $2J+1$ magnetic sublevels degenerate in the absence of magnetic fields, characterized by the magnetic quantum number $M$ ($M=-J,-J+1,\dots,J$). 
The magnetic field removes the degeneracy among the various sublevels (Zeeman effect), inducing energy splitting, that is,
\begin{equation*}
\Delta E=\nu_L g M\,,
\end{equation*}
where $\nu_\mathrm{L}$ is the Larmor frequency and $g$ is the Land\'e factor.
The spectral line takes into account the contribution of all the allowed transitions 
connecting an upper sublevel $(J_u M_u)$ and a lower sublevel $(J_\ell M_\ell)$.
Atomic polarization is neglected. 

\begin{figure}
\centering
\epsscale{1.1}
\includegraphics[width=.45\textwidth]{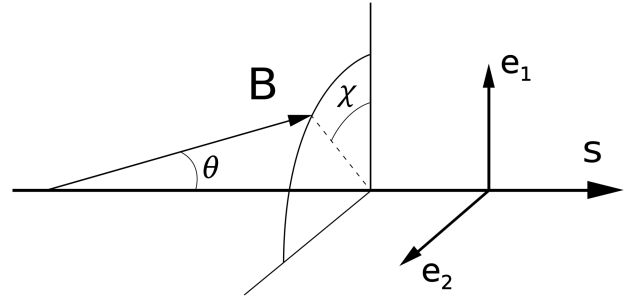}
\caption{Angles $\theta$ and $\chi$ specify the direction of the magnetic field $\mathbf B$ with respect to the coordinate system of the line-of-sight $s$. The Stokes component $Q$ is defined as the intensity difference of the linearly polarized light in the two orthogonal axes $\mathbf e_1$ and $\mathbf e_2$ in the plane perpendicular to the light beam.}
\label{fig:magnetic_field}
\end{figure}

Coming back to the matrix $\mathbf K$, the total absorption coefficient $\eta_I$  can be written as
\begin{equation*}
\eta_I=k_c + k_L \phi_I\,,
\end{equation*}
where $k_c$ is the local continuum absorption coefficient, $k_L$ is the (frequency integrated) line absorption coefficient, and $\phi_I$ is the intensity absorption profile.
Note that $\eta_I$ can always be assumed to be positive\footnote{Stimulated emission (which enters $k_L$) is capable of  producing 
an inversion of populations between two atomic levels. This could lead to a negative total absorption coefficient that yields an amplification of the radiation during the propagation. This phenomenon, which is at the basis of the devices such as lasers and masers, is completely negligible in solar applications and is not considered in this work.}.
The dichroism coefficients and the anomalous dispersion coefficients read
\begin{equation*}
\eta_i=k_L \phi_i\,,\quad\rho_i=k_L \psi_i\,,
\end{equation*}
respectively, where $i=Q,U,V$. When the orientation of the magnetic field $\mathbf B$ with respect to the line-of-sight is described with  
the inclination angle $\theta$ and the azimuth angle $\chi$ (as in Figure~\ref{fig:magnetic_field}), one has
\begin{align}
\phi_I &=\frac{1}{2}\left[\phi_0\sin^2\theta + \frac{\phi_{-1}+\phi_1}{2}\right](1+\cos^2\theta)\,,\nonumber\\
\phi_Q &=\frac{1}{2}\left[\phi_0 - \frac{\phi_{-1}+\phi_1}{2}\right]\sin^2\theta\cos 2\chi\,,\nonumber\\
\phi_U &=\frac{1}{2}\left[\phi_0 - \frac{\phi_{-1}+\phi_1}{2}\right]\sin^2\theta\sin 2\chi\,,\nonumber\\
\phi_V &=\frac{1}{2}\left[\phi_1 - \phi_{-1}\right]\cos\theta\,,\nonumber\\
\psi_Q &=\frac{1}{2}\left[\psi_0 - \frac{\psi_{-1}+\psi_1}{2}\right]\sin^2\theta\cos 2\chi\,,\nonumber\\
\psi_U &=\frac{1}{2}\left[\psi_0 - \frac{\psi_{-1}+\psi_1}{2}\right]\sin^2\theta\sin 2\chi\,,\nonumber\\
\psi_V &=\frac{1}{2}\left[\psi_1 - \psi_{-1}\right]\cos\theta\,.
\label{profiles}
\end{align}
In the observer's frame, the explicit expressions of the absorption profiles $\phi_q$ and the dispersion profiles $\psi_q$ $(q=-1,0,1)$ read, respectively,
\begin{align}
\phi_{q} &= \sum_{M_\ell, M_u}\! S_q^{J_\ell J_u}(M_\ell,M_u)\frac{1}{\sqrt{\pi}} H(\omega,a)\,,\label{eq:absorption_q}\\
\psi_{q} &= \sum_{M_\ell, M_u}\! S_q^{J_\ell J_u}(M_\ell,M_u)\frac{1}{\sqrt{\pi}} L(\omega,a)\,,\label{eq:dispersion_q}
\end{align}
where $S^{J_\ell J_u}_{q}(M_\ell, M_u)$ is the relative strength of the Zeeman component $q$ connecting the upper sublevel $(J_u M_u)$ and the lower sublevel $(J_\ell M_\ell)$. Using Wigner 3-$j$ symbols, its explicit expression is given by
\begin{equation*}
\label{eq:zeeman_strength}
 S_q^{J_\ell J_u}(M_\ell,M_u) = 3 
 \begin{pmatrix}
   J_u  & J_\ell & 1 \\
   -M_u & M_\ell & -q
 \end{pmatrix}^2\,.
\end{equation*}
The functions $H$ and $L$ appearing in
Formulas \eqref{eq:absorption_q} and \eqref{eq:dispersion_q}
correspond to the Voigt and Faraday-Voigt profiles defined by 
\begin{align*}
  H(\omega,a) &= \frac{a}{\pi}\int_{-\infty}^{\infty}{e}^{-x^2} \frac{1}{(\omega-x)^2 + a^2}\,{\rm d} x\,,\\
  L(\omega,a) &= \frac{1}{\pi}\int_{-\infty}^{\infty}{e}^{-x^2} \frac{\omega-x}{(\omega-x)^2 + a^2}\,{\rm d} x\,,
\end{align*}
respectively. Denoting with $g_u$ and $g_\ell$ the Land\'e factors associated to the upper and lower levels, respectively, the quantity $\omega$ is defined as
\begin{equation*}
    \omega = v\!-\!v_\mathrm{ A} + v_{B}(g_u M_u\! -\! g_\ell M_\ell) \,,
\end{equation*}
where the reduced frequency $v$ is defined by
\begin{equation*}
    v = \frac{\nu_0 - \nu}{\Delta \nu_\mathrm{ D}} \,,
\end{equation*}
with $\nu$ and $\nu_0$ being the frequency under consideration and line-center frequency, respectively. The Doppler width of the line $\Delta \nu_\mathrm{ D}$ is given by
\begin{equation*}
\Delta \nu_\mathrm{ D} = \frac{\nu_0 w_\mathrm{ T}}{c}\,,
\end{equation*}
where $w_{\rm T}$ denotes the random velocity of the atoms due to thermal and microturbulent motions, and $c$ is the speed of light. The quantity
\begin{equation*}
v_\mathrm{ A} = \frac{w_\mathrm{A}}{w_\mathrm{ T}}\,,
\end{equation*}
is the normalized frequency shift due to a bulk motion of velocity $w_\mathrm{ A}$ in the medium. The normalized Zeeman splitting $v_\mathrm{ B}$ is given by
\begin{equation*}
v_\mathrm{ B} = \frac{\nu_\mathrm{ L}}{\Delta \nu_\mathrm{ D}}\,.
\end{equation*}
The damping constant $a$ is given by
\begin{equation*}
    a = \frac{\Gamma}{\Delta \nu_\mathrm{D}} \,,
\end{equation*}
where $\Gamma$ takes into account the natural width of the line $\Gamma_n$ (due to the finite life-time of the upper and lower level) and the collisional width $\Gamma_c$ (due to collisions of the atom under consideration with other atoms and ions in the plasma) and it reads
\begin{equation*}
    \Gamma = \Gamma_n+\Gamma_c\,.
\end{equation*}
\subsection{Eigenvalues of the propagation matrix}\label{sec:eigK}
Let
\begin{equation*}
\boldsymbol{\eta}=(\eta_Q,\eta_U,\eta_V)^T\,,\quad\boldsymbol{\rho}=(\rho_Q,\rho_U,\rho_V)^T\,,
\end{equation*}
denote the dichroism and the anomalous dispersion vectors, respectively. The four eigenvalues of the propagation matrix $\mathbf K$ read \citep{landi_deglinnocenti+landolfi2004}
\begin{align}
\lambda^{(1)}&=\eta_I + \Lambda_+(\boldsymbol{\eta},\boldsymbol{\rho})\,,\nonumber\\
\lambda^{(2)}&=\eta_I - \Lambda_+(\boldsymbol{\eta},\boldsymbol{\rho})\,,\nonumber\\
\lambda^{(3)}&=\eta_I + {\rm i}\Lambda_-(\boldsymbol{\eta},\boldsymbol{\rho})\,,\nonumber\\
\lambda^{(4)}&=\eta_I - {\rm i}\Lambda_-(\boldsymbol{\eta},\boldsymbol{\rho})\,,
\label{eigenvalues}
\end{align}
where
\begin{align*}
\Lambda_+(\boldsymbol{\eta},\boldsymbol{\rho})&=\sqrt{\sqrt{(\eta^2-\rho^2)^2/4+(\boldsymbol{\eta}\cdot\boldsymbol{\rho})^2}+(\eta^2-\rho^2)/2}\,,\\
\Lambda_-(\boldsymbol{\eta},\boldsymbol{\rho})&=\sqrt{\sqrt{(\eta^2-\rho^2)^2/4+(\boldsymbol{\eta}\cdot\boldsymbol{\rho})^2}-(\eta^2-\rho^2)/2}\,,
\end{align*}
and
\begin{equation*}
\eta^2=\eta_Q^2+\eta_U^2+\eta_V^2\,,\quad\rho^2=\rho_Q^2+\rho_U^2+\rho_V^2\,.
\end{equation*}
The module of the dichroism vector satisfies 
\begin{equation}\label{eq:dichroism condition}
\eta\le\eta_I\,,
\end{equation}
but no similar relation holds for $\rho$. 
The comprehension of these expressions is facilitated by Table~\ref{tab:lambdas}, where the factors $\Lambda_+$ and $\Lambda_-$
are given for certain special cases.
Note that $\Lambda_+$ and $\Lambda_-$ do not depend on the azimuth angle $\chi$ of the magnetic field vector and they always assume real positive values limited by
\begin{equation}\label{eq:boundGamma}
0\le\Lambda_+\le\eta\,,\quad0\le\Lambda_-\le\rho\,.
\end{equation}
 %
The combination of conditions~\eqref{eq:dichroism condition} and~\eqref{eq:boundGamma} guarantees that the real part of the eigenvalues in Equation~\eqref{eigenvalues} is always positive.
Therefore, the spectral radius $r(\mathbf K)$ of the propagation matrix $\mathbf K$ satisfies
\begin{equation}\label{eq:spectralradius}
\eta_I \le\;  r(\mathbf K) = \eta_I \cdot\max\left\{ 1 + \Lambda_+/\eta_I, \sqrt{1+\Lambda_-^2/\eta_I^2}\right\} \,.
\end{equation}
Finally, knowing if the propagation matrix $\mathbf K$ is diagonalizable is relevant information, because stability analysis is notably simpler in this case.
If $\eta=0$ or $\rho=0$, the propagation matrix is normal (see Appendix~\ref{appendix:normal}) and, consequently, diagonalizable in $\mathbb{R}$.
If $\boldsymbol{\eta}\cdot\boldsymbol{\rho}\neq 0$, then both $\Lambda_+>0$
and $\Lambda_->0$. This implies that $\mathbf K$ has four distinct
eigenvalues and can be thus diagonalized in $\mathbb{C}$. On the other hand,
if $\boldsymbol{\eta}\perp\boldsymbol{\rho}$ and neither $\eta=0$ nor $\rho= 0$,
$\mathbf K$ may not be diagonalizable because its eigenvalues are not distinct (see
Table~\ref{tab:lambdas}).
\begin{table}
\caption{Factors $\Lambda_+$ and $\Lambda_-$ for different values of $\eta$ and $\rho$}
\centering
\begin{tabular}{|c|c|c|}
\hline
\emph{Special cases} & $\Lambda_+$ & $\Lambda_-$\\ 
\hline
$\eta=\rho=0$ & 0  & 0\\ 
$\rho=0$ & $\eta$ & 0\\ 
$\eta=0$ & 0 & $\rho$\\ 
$\boldsymbol{\eta}\parallel\boldsymbol{\rho}$ & $\eta$ & $\rho$\\ 
$\boldsymbol{\eta}\perp\boldsymbol{\rho}$ and $\eta=\rho$ & 0 & 0\\
$\boldsymbol{\eta}\perp\boldsymbol{\rho}$ and $\eta>\rho$ & $\sqrt{\eta^2-\rho^2}$ & 0\\
$\boldsymbol{\eta}\perp\boldsymbol{\rho}$ and $\rho>\eta$ & 0 & $\sqrt{\rho^2-\eta^2}$\\
\hline
\end{tabular}
\label{tab:lambdas}
\end{table}
 %
\section{Stability analysis}\label{sec:stability}
Performing stability analysis of numerical methods for ODEs is often quite involved.
A gentle introduction
to stability analysis of numerical methods for ODEs can be found in \citet{higham1993}.

This section is dedicated to the study of the stability properties of Runge-Kutta methods applied to
Equation~\eqref{eq:RTE}.
In this equation, the Stokes vector $\mathbf I$ is the only quantity that can
propagate or amplify errors introduced in previous steps.
Consequently, the emission term $\boldsymbol{\epsilon}$ can be omitted in the stability analysis,
because it does not explicitly depend on $\mathbf I$.

Moreover, Equation~\eqref{eq:RTE} is linear in the variable $\mathbf{I}$ and the propagation matrix
$\mathbf{K}$ depends on the space variable $s$.
In this case,  
it is common to analyze the dynamics of the system 
assuming that $\mathbf{K}$ is constant around each position $s_0$ of interest.
Denoting by $\mathbf{A}=-\mathbf K (s_0)$ the propagation matrix with ``frozen'' coefficients,
one easily performs the stability analysis on the simpler initial value problem (IVP)
\begin{equation}
\begin{aligned}
\mathbf{y}'(t)= \mathbf{A}\mathbf{y}(t)\,,\quad
\mathbf{y}(0)=\mathbf{y}_0\,.
\label{IVP1}
\end{aligned} 
\end{equation}
The remainder of this section is structured as follows: Section~\ref{subsec:RKdiag} presents the stability analysis further assuming that the ``frozen'' matrix $\mathbf{A}$ is diagonalizable. This particular case is notably simpler, because the linear system of ODEs is reduced to a set of scalar problems via diagonalization.
Section~\ref{subsec:RKnonnormal} analyzes the case of a more general (non-diagonalizable) ``frozen'' matrix.
Finally, Section~\ref{subsec:Kvariations} addresses the limits due to the ``frozen'' matrix assumption, by investigating
how spatial variations in matrix $\mathbf{A}$ affect the stability of numerical methods.
\subsection{Reduction to the scalar case}\label{subsec:RKdiag}
A matrix $\mathbf{A}$ is called diagonalizable if there is an invertible matrix $\mathbf{U}$ such that
\begin{equation*}
\mathbf{A} = \mathbf{U}^{-1}\mathbf{D}\mathbf{U}\,, 
\end{equation*}
where $\mathbf{D}$ is a diagonal matrix whose entries are the eigenvalues of $\mathbf{A}$.
From Equation~\eqref{IVP1}, it is easy to see that $\mathbf x = \mathbf U\mathbf y$ satisfies
\begin{equation}
\mathbf{x}'(t)= \mathbf{D}\mathbf{x}(t)\,,\quad 
\mathbf{x}(0)=\mathbf{U}\mathbf{y}_0\,.
\label{eq:diagODE}
\end{equation}
Runge-Kutta methods are affine covariant.
This means that the very same approximation
of $\mathbf y$ that is obtained by applying a Runge-Kutta method to the IVP~\eqref{IVP1}
can be computed applying the Runge-Kutta method to the IVP~\eqref{eq:diagODE} first and
multiplying the result with the matrix $\mathbf U^{-1}$ at the end.
For this reason, the IVP \eqref{IVP1} can be replaced by the IVP~\eqref{eq:diagODE},
and since the latter is a system of decoupled differential equations,
it is sufficient to consider the scalar case
\begin{equation}\label{eq:Dahlquist}
x'(t) = \lambda x(t)\,, \quad x(0) = x_0\,,
\end{equation}
where $\lambda$ represents any of the eigenvalues of $\mathbf A$. 

The solution of the IVP \eqref{eq:Dahlquist} is given by
\begin{equation*}
x(t)= x_0e^{\lambda t}\,.
\end{equation*}
When $\operatorname{Re}(\lambda)< 0$, $x(t)$ converges to zero as $t \rightarrow \infty$.
The imaginary part of $\lambda$ only introduces an oscillatory behavior of the solution.

Let $\{t_{k}\}$ be a {discrete grid}, and let $x_k \approx x(t_k)$ be a numerical  
solution computed with a Runge-Kutta method. Then, $x_k$ and $x_{k+1}$ satisfy
\begin{equation}\label{eq:stabstep}
x_{k+1}=\phi(\lambda\Delta t)x_k\,,
\end{equation}
where $\Delta t= t_{k+1}-t_k$,  
and $\phi$ is the stability function of the numerical method \citep{frank2008}.  

A numerical solution of an IVP is said to be asymptotically stable if the sequence $\{x_{k}\}$
converges to zero for $k \rightarrow \infty$.
Intuitively, this guarantees that any perturbation in the solution is attenuated with the recursive numerical integration. 
In light of Equation~\eqref{eq:stabstep}, asymptotic stability is equivalent to
\begin{equation}
\vert\phi(\lambda\Delta t)\vert<1\,.
\label{stability_condition}
\end{equation}
The stability of a numerical solution is therefore related to both the step-size $\Delta t$ and the eigenvalue $\lambda$.
More precisely, it depends on the product $\lambda\Delta t$. The stability region $S$ of a Runge-Kutta method is defined as the set of complex values $z=\lambda\Delta t$ for which Equation~\eqref{stability_condition} is satisfied, that is,
\begin{equation*}
S = \{ z\in \mathbb{C}: \vert\phi(z)\vert<1\}\,. 
\end{equation*}
To give an example, the stability function of the explicit Runge-Kutta 4 method is given by
 %
\begin{equation*}
\phi_{\text{\tiny RK4}}(z)=1+z+\frac{z^2}{2}+\frac{z^3}{6}+\frac{z^4}{24}\,,
\end{equation*}
and it is displayed in yellow in Figure~\ref{pseudospectrum}.

If $\operatorname{Re}(\lambda)< 0$, asymptotic stability is
guaranteed when
$\lambda\Delta t$ lies inside the stability region of the numerical method.
For the particular case of Equation~\eqref{eq:RTE},
Section~\ref{sec:eigK} shows that the real part of the eigenvalues of 
the propagation operator $-\mathbf{K}$ is always
nonpositive. Since its eigenvalues are known explicitly,
Equation~\eqref{eq:spectralradius} can be used to derive a sharp upper bound on the step-size $\Delta s$ that ensures asymptotic stability of the numerical solution.

If the Runge-Kutta method is consistent, complex numbers $z$ with negative real part and sufficiently small absolute value 
lie in the stability region $S$. Therefore, for consistent methods, 
instabilities can be prevented by choosing a sufficiently small step-size $\Delta t$.
However, the downside of small step-sizes is that, for a fixed  
integration interval, the number of integration steps increases.

To overcome the need of choosing very small step-sizes,
the stability region of the numerical method employed should be as large as possible.
In particular, to ensure that the numerical solution remains asymptotically stable
independently of the choice of $\Delta t$,  the stability region
should comprise the complex left half-plane $\mathbb{C}^{-}$.
Runge-Kutta methods that satisfy this condition are called
$A$-stable, and
one of the simplest $A$-stable Runge-Kutta methods is the (implicit) trapezoidal method.

$A$-stability guarantees that the numerical solution is stable if $\mathrm{Re}(\lambda) <0$. 
However, if $\mathrm{Re}(\lambda\Delta t)$ is a large negative value,
$A$-stability may not be sufficient to replicate the exponential decay of 
the sequence {$\{x(k \Delta t)\}$}, because $A$-stability does not guarantee that 
\begin{equation}\label{eq:Lstable}
\lim_{\mathrm{Re}(z)\to -\infty} \phi(z) = 0\,.
\end{equation}
For instance, the stability function of the trapezoidal method satisfies
$\lim_{\mathrm{Re}(z)\to -\infty} \phi(z) = -1$. In this case, the numerical solution $\{x_k\}$
will still converge to zero, but the decay becomes arbitrarily slow as
$\mathrm{Re}(\lambda \Delta t)\to -\infty$.
 %
$A$-stable Runge-Kutta methods that further satisfy condition~\eqref{eq:Lstable}
are called $L$-stable,
 %
and they correctly replicate
exponential attenuations even when the step-size is large. The simplest $L$-stable Runge-Kutta method is the implicit (or backward) Euler scheme.

Runge-Kutta methods are also classified into explicit and implicit methods.
A Runge-Kutta method that does not require solving a system of equations
to update the solution is called explicit; otherwise, it is
called implicit. Clearly, explicit methods are computationally less expensive.
However, explicit Runge-Kutta methods cannot be $A$-stable, and therefore also not $L$-stable.  
Diagonally implicit Runge-Kutta methods \citep[e.g.,][]{kennedy2016diagonally}
offer a good compromise between stability, order of accuracy, and computational
complexity. For instance, second-order $L$-stable diagonally implicit Runge-Kutta methods are available. When applied to linear ODEs like Equation~\eqref{eq:RTE},
their computational cost is particularly competitive
because it grows only linearly with respect to the number of Runge-Kutta stages.
However, the same computational cost may grow as $d^3$, where
$d$ is the dimension of the ODE. Note that $d=4$ in Equation~\eqref{eq:RTE}.
%
 %
\subsection{Analysis of pseudospectra}\label{subsec:RKnonnormal}
The stability analysis presented in the previous section
hinges on assuming that the matrix $\mathbf{A}$
in Equation~\eqref{IVP1} is diagonalizable and only considers the scalar IVP~\eqref{eq:Dahlquist} with $\lambda$ representing any of the eigenvalues of $\mathbf A$.
 %
However, Section~\ref{sec:eigK} shows that, in general,
the propagation matrix $\mathbf K$ may not be diagonalizable.

Instead of employing eigenvalues, \citet{higham1993} suggest to 
perform stability analyses focusing on pseudospectra
(more details on pseudospectra are given in Appendix~\ref{appendix:B}).
 %
They point out that  
the  
analysis based on eigenvalues can lead to too liberal conditions for the absence of stiffness,
because 
eigenvalues describe the asymptotic
behavior only, whereas instability and stiffness are transient phenomena that depend
on how the effects compound over few integration steps.
For instance, if the spectrum $\sigma(\mathbf{A})\subset \mathbb{C}^-$, then
the solution $\mathbf{y}(t)$ to the IVP~\eqref{IVP1} satisfies $\lim_{t\to\infty}\Vert \mathbf{y}(t) \Vert = 0$. However,
the decay of $\Vert \mathbf{y}(t) \Vert$ may not be monotone.
\citet{higham1993} conclude that numerical instability around $t$ in Equation~\eqref{IVP1} occurs when the pseudospectra of the frozen coefficient matrix
$\Delta t\mathbf A$ fail to fit within the stability region $S$ of the numerical method.
This alternative stability analysis is particularly insightful when
the pseudospectra of $\mathbf A$ are highly dispersed.
Unfortunately, computing pseudospectra is a computationally demanding task
that is not affordable in  
real-time computations.
Nevertheless, one can rely on two generic observations.
First, if $\mathbf A$ is normal, its pseudospectrum
is tightly clustered around the spectrum and the difference between transient and asymptotic
stability behaviors is irrelevant.
Second, pseudospectra tend to be particularly dispersed
if $\mathbf A$ is both non-normal and ``close'' to a non-diagonalizable matrix.



Appendix~\ref{appendix:normal} shows by direct calculation that the propagation matrix $\mathbf K$ is normal if and only if
\begin{equation*}\label{eq:normalitycondition}
\boldsymbol{\eta}\times \boldsymbol{\rho}= \boldsymbol{0}\,.
\end{equation*}
In particular,  
if $\boldsymbol{\eta}\perp\boldsymbol{\rho}$ and neither $\eta=0$ nor $\rho= 0$,
$\mathbf{K}$ could be both non-normal and non-diagonalizable.
Moreover, numerical tests show that an additional requirement to produce largely dispersed pseudospectra is given by
$\eta\approx\rho\gg \eta_I$. This empirical condition assures that the four eigenvalues are degenerate (see Table~\ref{tab:lambdas}).

However, the entries of the propagation matrix $\mathbf K$ satisfy
the dichroism condition~\eqref{eq:dichroism condition}.
This condition guarantees that the ``empirical condition'' above is never satisfied and it prevents the pseudospectra from being dispersed.
For this reason,
the diagonalization step
performed in Section~\ref{subsec:RKdiag} does not pose relevant problems in the stability analysis.

\begin{figure*}
\centering
\includegraphics[width=1.\textwidth]{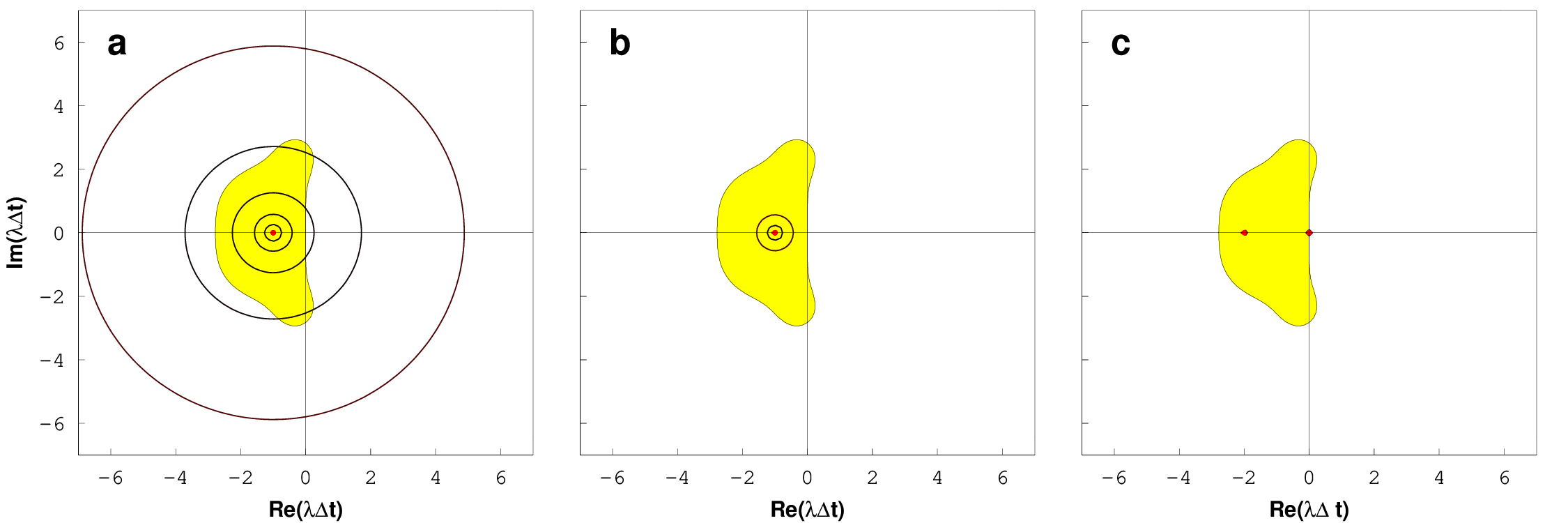}
  \caption{Pseudospectra for the matrices {\bf a)} $\mathbf K_1$, {\bf b)} $\mathbf K_2$, and {\bf c)} $\mathbf K_3$ given in Equation~\eqref{eq:matrices}.
  The black circles (which are almost invisible in {\bf c)}) are the boundaries of the $\epsilon$-pseudospectrum $\Lambda_{\epsilon}$ for $\epsilon=10^{-1},10^{-2},10^{-3},\dots$, where the outermost curve corresponds to $\epsilon=10^{-1}$. The red dots represent the eigenvalues of the matrix, which are four times degenerate in {\bf a)} and {\bf b)} and twice degenerate in {\bf c)}.   In yellow, the stability region for the explicit Runge-Kutta 4 method.}
\label{pseudospectrum}
\end{figure*}

Some numerical evidence is presented in Figure~\ref{pseudospectrum},
which displays the pseduospectra (for $\Delta s =1$) of the three matrices
{\small
\begin{gather}
 \nonumber \mathbf K_1=\begin{pmatrix}
      1 & 30 & 10 & 0 \\[5pt]
      30 & 1 & 0 & 30 \\[5pt]
      10 & 0 & 1 & 10\\[5pt]
      0 & -30 & -10 & 1 
  \end{pmatrix}\,,\quad
  \mathbf K_2=\begin{pmatrix}
      1 & \frac{8}{10} & \frac{3}{10} & 0 \\[5pt]
      \frac{8}{10} & 1 & 0 & -\frac{8}{10} \\[5pt]
      \frac{3}{10} & 0 & 1 & -\frac{3}{10} \\[5pt]
      0 & \frac{8}{10} & \frac{3}{10} & 1 
  \end{pmatrix}\,,\\
  \mathbf K_3=\begin{pmatrix}
      1 & \frac{\sqrt{2}}{2} & 0 & \frac{\sqrt{2}}{2} \\[5pt]
      \frac{\sqrt{2}}{2} & 1 & -20 & 0 \\[5pt]
      0 & 20 & 1 & -20 \\[5pt]
      \frac{\sqrt{2}}{2} & 0 & 20 & 1 
  \end{pmatrix}\,.
  \label{eq:matrices}
\end{gather}}
 %
 %
The matrices $\mathbf K_1$ and $\mathbf K_2$ satisfy $\boldsymbol{\eta}\perp\boldsymbol{\rho}$ and $\eta=\rho\ne0$,
showing dispersed pseudospectra in Figures~\ref{pseudospectrum}a and~\ref{pseudospectrum}b, respectively.
Due to its (unphysically) large $\eta$ and $\rho$, $\mathbf K_1$ presents
a remarkably scattered pseudospectrum,
while $\mathbf K_2$, which satisfies condition~\eqref{eq:dichroism condition}, shows a tighter pseudospectrum.
Figure~\ref{pseudospectrum}c shows that the pseudospectrum of the matrix $\mathbf K_3$, which does not satisfy the condition $\eta\approx\rho$,
is not dispersed. 
Experiments performed for different values of $\Delta s$ lead to similar results.
 %
\subsection{Variation of eigenvalues along the integration path}\label{subsec:Kvariations}
The stability analyses presented in Sections \ref{subsec:RKdiag} and
\ref{subsec:RKnonnormal} neglect the dependence of the
propagation matrix $\mathbf K$ on the spatial variable $s$.
The following example shows that, in principle, stability issues may
arise even when numerical integrations 
are based on $A$-stable methods.
\citet{janett2017a} give graphical illustrations of this phenomenon in terms of modifications 
of the stability region of the trapezoidal method.

Consider the scalar test case
\begin{equation}\label{eq:testcase}
y' = \lambda(t) y\,, \quad y(t_0)=y_0\,,
\end{equation}
where $\lambda(t)$ is a real function.
The numerical approximation $y_1$ of $y(t_0 + \Delta t)$
computed with the trapezoidal method, which is $A$-stable, reads
\begin{equation}\label{eq:trapezoidal}
y_1 = \left(\frac{1+\lambda(t_0)\Delta t/2}{1-\lambda(t_0+\Delta t)\Delta t/2}\right) y_0\,. 
\end{equation}
If $\lambda(t)$ is negative for $t\in[t_0, t_0+\Delta t]$, then
$\vert y(t_0+\Delta t)\vert < \vert y_0\vert$. For the numerical method to be stable, it is necessary that $\vert y_1 \vert< \vert y_0\vert$ as
well. However, this is not always guaranteed
if $-\lambda(t_0)\Delta t > 2$. For instance, if
$\Delta t = -12/\lambda(t_0)$ and
$\lambda(t_0+\Delta t)=\lambda(t_0)/2 < 0$,
then $y_1 = (-5/4)y_0$. This means that stability can be lost for sufficiently large variations in $\lambda(t)$ and $\Delta t$ big enough, and this despite the trapezoidal method being $A$-stable.
This example shows that large variations in the coefficient $\lambda$ affect
the stability of the trapezoidal scheme.
In particular, the stability region depends on the values
$\lambda(t_0) \Delta t$ and $\lambda(t_0 + \Delta t) \Delta t$.

More generally, the stability region of a Runge-Kutta method
depends on how much the quantity
$\lambda(t_0 + x\Delta t) \Delta t$ varies for $x\in[0,1]$.
Let us assume that the function $\lambda(t)$
is continuous\footnote{This assumption is
required by the Picard-Lindel\"of Theorem to ensure that the IVP \eqref{eq:testcase}
has a solution, and that this solution is unique.}. Continuous dependence on $t$
implies that variations of $\lambda(t_0 + x\Delta t)$, for $x\in[0,1]$,
can be controlled by choosing a sufficiently small $\Delta t$.
In turn, this implies that the smaller $\Delta t$, 
the tighter the bound on the variations of the quantity $\lambda(t_0 + x\Delta t) \Delta t$,
so that numerical stability can be recovered.

For the sake of completeness, one should mention that in the nonscalar case there are examples of non-constant 
matrices  $\mathbf{B}(t)$ that satisfy
$\sigma(\mathbf{B}(t))\subset \mathbb{C}^-$ for every $t$, and for which
the solution to the IVP
\begin{equation*}
\mathbf{y}'(t)= \mathbf{B}(t)\mathbf{y}(t)\,,\quad
\mathbf{y}(0)=\mathbf{y}_0\,.
\end{equation*}
satisfies $\lim_{t\to\infty}\Vert \mathbf{y}(t) \Vert = \infty$ \citep{JoRo08}.
In general, this happens when the matrix is non-normal and is related to its
pseudospectra being dispersed. However, in light of the discussion presented in
Section~\ref{subsec:RKnonnormal}, it is unlikely that Equation~\eqref{eq:RTE}
supports this kind of unstable solution.
 %

\section{Conversion to optical depth}\label{sec:conversion}
 %
 %
To reduce variations of the propagation matrix $\mathbf K$ along the ray path,
several authors \citep[e.g.,][]{rees+al1989,delacruz_rodriguez+piskunov2013} suggest
to rewrite Equation~\eqref{eq:RTE} in terms of the optical depth $\tau$. This idea
was first exploited to devise numerical schemes for the transfer
of unpolarized light, providing significant stability enhancements (see Appendix~\ref{sec:unpolarizedlight}).

To transplant Equation~\eqref{eq:RTE} to the optical depth regime,
one can consider the map $g:[s_0, s_{\rm f}]\to [\tau_0, \tau_{\rm f}] \subset \mathbb{R}^+$,
defined as the solution of the IVP\footnote{In the literature, $g$ is usually
defined as the solution of
$g'(s) = -\eta_I(s)$. However, the negative sign induces an unnecessary and
possibly confusing change in the integration direction.}
\begin{equation}\label{eq:conversion}
g'(s) = \eta_I(s)\,, \quad g(s_0) = \tau_0\,.
\end{equation}
where\footnote{The subscript ``${\rm f}$'' stands for ``final''.} $\tau_{\rm f} = g(s_{\rm f})$.
Since $\eta_I>0$, $g$ is strictly monotone increasing and thus a (differentiable)
bijection from $[s_0, s_{\rm f}]$ to $[\tau_0, \tau_{\rm f}]$.
Let $\mathbf{Z}:[\tau_0, \tau_{\rm f}]\to \mathbb{R}^4$ be defined by
\begin{equation*}
\mathbf{I}(s) = \mathbf{Z}(g(s))\quad \text{for every } s \text{ in } [s_0, s_{\rm f}] \,.
\end{equation*}
On the one hand, by direct differentiation,
\begin{align*}
\mathbf{I}(s)'  = \mathbf{Z}'(g(s))\cdot g'(s) =  \eta_I(s)\mathbf{Z}'(g(s))\,.
\end{align*}
On the other hand, from Equation~\eqref{eq:RTE},
\begin{align*}
\mathbf{I}(s)'  = -\mathbf{K}(s)\mathbf{Z}(g(s))+\boldsymbol{\epsilon}(s)\,.
\end{align*}
By combining the two equations above, one obtains 
\begin{equation}\label{eq:RTEod}
\begin{split}
\frac{\rm d}{\rm d\tau}\mathbf{Z}(\tau) &=
-\frac{\mathbf{K}(g^{-1}(\tau))}{\eta_I(g^{-1}(\tau))}\mathbf{Z}(\tau)
+ \frac{\boldsymbol{\epsilon}(g^{-1}(\tau))}{\eta_I(g^{-1}(\tau))}\\
&=-\tilde{\mathbf K}(g^{-1}(\tau))\mathbf{Z}(\tau)
+ \tilde{\boldsymbol{\epsilon}}(g^{-1}(\tau))\,,
\end{split}
\end{equation}\noindent
which is formally equivalent to Equation~\eqref{eq:RTE}.

The matrix $\tilde{\mathbf K}=\mathbf K/\eta_I$ satisfies
\begin{equation*}
\label{eq:line_propagation_matrix}
  \tilde{\mathbf K} =
      \begin{pmatrix}
      1 &  h_Q &  h_U & h_V  \\
      h_Q &  1 &  r_V & -r_U \\
      h_U & -r_V &  1 & r_Q  \\
      h_V &  r_U & -r_Q & 1 
      \end{pmatrix}\,
\end{equation*}
where, for $i=Q,U,V$,
\begin{equation*}
h_i=\frac{\eta_i}{\eta_I}=\frac{k_L \phi_i}{k_c+k_L \phi_I}\,,\quad r_i=\frac{\rho_i}{\eta_I}=\frac{k_L \psi_i}{k_c+k_L \phi_I}\,,
\label{K_tilde_coefficients}
\end{equation*}
and the modified emission vector is given by $\tilde{\boldsymbol{\epsilon}}=\boldsymbol{\epsilon}/\eta_I$.

The eigenvalues of the propagation matrix $\tilde{\mathbf K}$
can be directly expressed in terms of the eigenvalues of ${\mathbf K}$, namely
\begin{equation}
  \tilde{\lambda}^{(i)}(\tau)=\frac{\lambda^{(i)}(g^{-1}(\tau))}{\eta_I(g^{-1}(\tau))}\,,
  \quad\text{ for } i=1,2,3,4\,.
\label{eigenvalues_tau}
\end{equation}
 %
 %
 %
In light of Equations~\eqref{eigenvalues}--\eqref{eq:boundGamma},
the spectral radius $r(\tilde{\mathbf K})$ of $\tilde{\mathbf K}$ satisfies
\begin{align*} 
1\leq r(\tilde{\mathbf K}) &= \max\left\{ 1 + \Lambda_+/\eta_I, \sqrt{1+\Lambda_-^2/\eta_I^2}\right\}\\
&\leq \max\left\{ 2, \sqrt{1+\rho^2/\eta_I^2}\right\}\,,
\end{align*}
and the real part of the eigenvalues of the propagation operator $-\tilde{\mathbf{K}}$ is always nonpositive.

 %
 %

The conversion to optical depth given by Equation~\eqref{eq:conversion} freezes the diagonal elements of $\tilde{\mathbf{K}}$ to 1.
The variations of its off-diagonal coefficients can be
estimated in the following two limiting cases.

Case 1: the continuum absorption is much larger than line processes, i.e., $k_c\gg k_L\phi_I$, and $k_c\gg \eta_i,\rho_i$ for $i=Q,U,V$.
In this case, the off-diagonal coefficients of $\tilde{\mathbf{K}}$, and in turn the absolute value of their variations, tend to zero.
Equations~\eqref{eigenvalues_tau} and~\eqref{eigenvalues} imply that the eigenvalues of $\tilde{\mathbf{K}}$ are close to 1.

Case 2: the line absorption dominates over the continuum absorption, i.e., $k_L\phi_I\gg k_c$.
In this case, the variation along the ray path of the off-diagonal coefficients of $\tilde{\mathbf{K}}$ is basically independent of $k_c$ and $k_L$.
Equations~\eqref{eigenvalues_tau} and~\eqref{eigenvalues} imply that the eigenvalues of $\tilde{\mathbf{K}}$
are also basically independent of $k_c$ and $k_L$ and their variations are only due to variations in the profiles given by Equation~\eqref{profiles}.

However, if the conditions of the two cases presented above are not met,
it is not straightforward to infer conclusions on the values of the off-diagonal
entries of $\tilde{\mathbf{K}}$, and strong variations in the propagation matrix may still be present
(in particular, due to the dependence of the coefficients given by Equation~\eqref{profiles} on variations of the magnetic field and of the bulk motions). 

In conclusion, the conversion to optical depth usually
reduces the amount of fluctuations of the propagation operator $-\mathbf K$ along
the ray path, but this is not guaranteed in general. 
\section{Numerical conversion to optical depth}\label{sec:numericalconversion}
To apply a numerical scheme to Equation~\eqref{eq:RTEod} it is necessary to have a certain
knowledge of the function $g$. 
From Equation~\eqref{eq:conversion}, one has
\begin{equation}
g(s) = \tau_0 + \int_{s_0}^s \eta_I(x)\,{\rm d}x\,,
\label{eq:conversion_opt_depth}
\end{equation}
and numerical approximations of $g$ can be obtained by replacing the integral  
with a numerical quadrature. It is absolutely
crucial that this numerical approximation is strictly monotone increasing
because one needs to access the values of its inverse $g^{-1}$.
Replacing $g$ with a numerical approximation could negatively affect the order
of the method employed to solve the IVP~\eqref{eq:RTEod}.
 %
\citet{janett2017a} explain that high-order solvers require a corresponding high-order numerical approximation of the integral in Equation~\eqref{eq:conversion_opt_depth}.
A very common  
approach to devise numerical quadrature rules is
to replace integrands with interpolants that are successively integrated exactly.
For instance, Gauss, Radau, Hermite, and Clenshaw-Curtis quadratures are based on
this idea. 

Here are some more concrete examples.
The trapezoidal rule applied to Equation~\eqref{eq:conversion_opt_depth} reads
\begin{equation*}
g(s) \approx \tau_0 + (s-s_0) \frac{\eta_I(s_0)+\eta_I(s)}{2}\,.
\end{equation*}
This quadrature
is based on a linear interpolation of $\eta_I$ through the points $\{s_0, s\}$ and is second-order accurate.
Higher-order monotone quadrature schemes can be obtained by replacing
linear interpolation with higher-order
monotone interpolants.

A concrete high-order  
example is given by
the monotone cubic Hermite quadrature, that, applied to Equation~\eqref{eq:conversion_opt_depth}, leads to 
{ 
\begin{equation*}
g(s) \approx \tau_0 + (s-s_0)\frac{\eta_I(s_0)+\eta_I(s)}{2}+(s-s_0)^2\frac{\tilde\eta_I'(s_0)-\tilde\eta_I'(s)}{12}\,,
\label{hermite_quadrature}
\end{equation*}}\noindent
where $\tilde\eta_I'$ are suitable numerical approximations (of the first derivative $\eta_I'$)
that guarantee monotonicity.
The approximation above is fourth-order accurate
provided that the approximation $\tilde\eta_I'\approx\eta_I'$ is at least of second order
\citep{dougherty1989}.
The approximation $\tilde\eta_I'$ described by \citet{steffen1990} satisfies both conditions,
whereas the one described by \citet{fritsch1984} guarantees monotonicity, but it is
second-order accurate on uniform grids only
(it drops to first-order on non-uniform grids).

Hermite interpolation is not the only option for higher-order monotone interpolation schemes.
For instance, both \citet{auer2003} and \citet{delacruz_rodriguez+piskunov2013} prefer to employ monotonic quadratic B\'ezier splines.
However, the high-order convergence of B{\'e}zier interpolations is
achieved only when  
the B{\'e}zier interpolants are forced to be
identical to the corresponding degree Hermite interpolants,
which do not guarantee monotonicity.


Finally, when the atmospheric model is exponentially stratified along the ray path, \citet{mihalas1978} suggests to replace $\eta_I$ with the exponential
function\footnote{If $\eta_I$
is known at the grid points $\{s_i\}_{i=1}^N$, it is natural to employ the piecewise
exponential model by adapting the parameter $\alpha$
to every interval $[s_i, s_{i+1}]$.}
\begin{equation*}
\eta_I(s_0) e^{(x-s_0)/\alpha}\,,\text{ for }x\in[s_0,s]\,,
\end{equation*}
with
\begin{equation*}
\alpha = \frac{(s-s_0)}{\log \eta_I(s)-\log \eta_I(s_0)}\,.
\end{equation*}
After such a substitution, Equation~\eqref{eq:conversion_opt_depth}
can be integrated exactly. The resulting map reads
\begin{equation*}
g_M(s) = \tau_0 + (s-s_0)\frac{\eta_I(s)-\eta_I(s_0)}{\log \eta_I(s)-\log \eta_I(s_0)}\,.
\end{equation*}
The error introduced by this substitution is bounded by
\begin{align*}
\vert g_M(s) - g(s)\vert
&\leq \int_{s_0}^s \vert \eta_I(s) - \eta_I(s_0) e^{(x-s_0)/\alpha}\vert
\,{\rm d}x\,,\\
&\leq (s-s_0) \max_{x\in[s_0, s]}\vert \eta_I(s) - \eta_I(s_0) e^{(x-s_0)/\alpha}\vert\,,
\end{align*}
and its accuracy clearly depends on the suitability of the exponential modeling.
\section{Pragmatic Formal solver}\label{sec:bestsolver}
In practical applications, the propagation matrix $\mathbf{K}$ and the emission
vector $\boldsymbol\epsilon$ are known only at a discrete set of usually nonequidistant grid points $\{s_i\}_{i=1}^N$.
In such an instance, one aims at computing
a numerical solution of Equation~\eqref{eq:RTE} that is first of all physically meaningful
(i.e., stable) and that, secondly, has a good ratio between accuracy and
computational cost.
Ideally, the numerical method should rely as much as possible
on the provided values of  $\mathbf{K}$ and $\boldsymbol\epsilon$,  
although these functions can be evaluated at other depths $s$ via interpolation, if necessary.
This interpolation must be sufficiently accurate to preserve
the order of convergence of the numerical scheme used for numerical integration \citep{janett2017b}.

Since stiffness is a transient behavior,
the analysis presented in the previous sections suggests to
consider each interval $[s_i, s_{i+1}]$ at a time and sequentially.
In each interval, depending on the cell width $\Delta s$ and on the magnitude of the
eigenvalues of $\mathbf{K}$ at $s_i$ and $s_{i+1}$, the approximation of $\mathbf{I}(s_{i+1})$ is computed with either
an explicit method $\Psie$ (which is computationally inexpensive)
or an $A$-stable method $\Psia$,  
or an $L$-stable method $\Psil$. Preferably, 
the methods $\Psie$ and $\Psia$ should be of the same order, while the method
$\Psil$ could be of lower order, {because large attenuations usually prevent any propagation of information.}

The following criteria can help in choosing between
$\Psie$, $\Psia$, or $\Psil$: (i) if the absolute value 
of the real part of the eigenvalues multiplied by the cell width is large,  
one should use $\Psil$ to guarantee the correct exponential attenuation of the Stokes vector.
Otherwise, (ii) the method $\Psie$ is used whenever stable (to reduce computational cost),
and (iii) if $\Psie$ is not stable, one uses $\Psia$
with the optional conversion to optical depth if $\Psia$
loses stability due to the variations of the eigenvalues in the interval $[s_i, s_{i+1}]$.

For example, this strategy can be implemented using 
Heun's method
(which is also known as the explicit trapezoidal rule and has order 2)
as $\Psie$, the implicit trapezoidal rule (which also has order 2) as $\Psia$ ,
and the implicit Euler method (which has order 1)   as $\Psil$ .
These methods employ $\mathbf{K}$ and $\boldsymbol{\epsilon}$ at grid points only, avoiding
the use of interpolated  
off-grid points' quantities.
Computing the eigenvalues of $\mathbf{K}$
at a point $s$ is roughly one-third as expensive\footnote{This fraction decreases if Heun's method is replaced
by a higher-order explicit Runge-Kutta scheme, because the latter inevitably requires the computation
of more stages.} as one step of  $\Psie$, whereas $\Psia$ is
roughly twice as expensive as $\Psie$.
The implicit Euler method is less expensive than $\Psia$, but more than
$\Psie$.
A second-order $L$-stable method would be at least as expensive as
$\Psia$, but since $L$-stability is only required when large
exponential attenuations are present,  
one can opt for a lower-order scheme.

To assess the stability of Heun's method $\Psie$, one should verify that
\begin{equation*}
\vert \phi_{\Psie}\vert = \left\vert
1 + \Delta s\frac{\lambda(s_i)+  \lambda(s_{i+1})
+\Delta s\lambda(s_i)\lambda(s_{i+1})}{2}
\right\vert <1\,.
\end{equation*}
However, it is worth distinguishing the cases when
$\vert \phi_{\Psie}\vert$ is close to 1: if $\lambda\Delta s$ is close
to 0, $\Psie$ can be trusted; however, if 
$\lambda\Delta s$ is close to the boundary of the stability domain away from 0,
it is advisable to switch to $\Psia$, because
$\Psie$ may suffer from instability.
To verify the stability of $\Psia$ and decide whether to
opt for the conversion to optical depth,
one can repeat the same argument used for 
$\Psie$ but using Formula~\eqref{eq:trapezoidal} instead of $\phi_{\Psie}$.

\begin{figure*}
\centering
\includegraphics[width=0.24\textwidth, height=4cm]{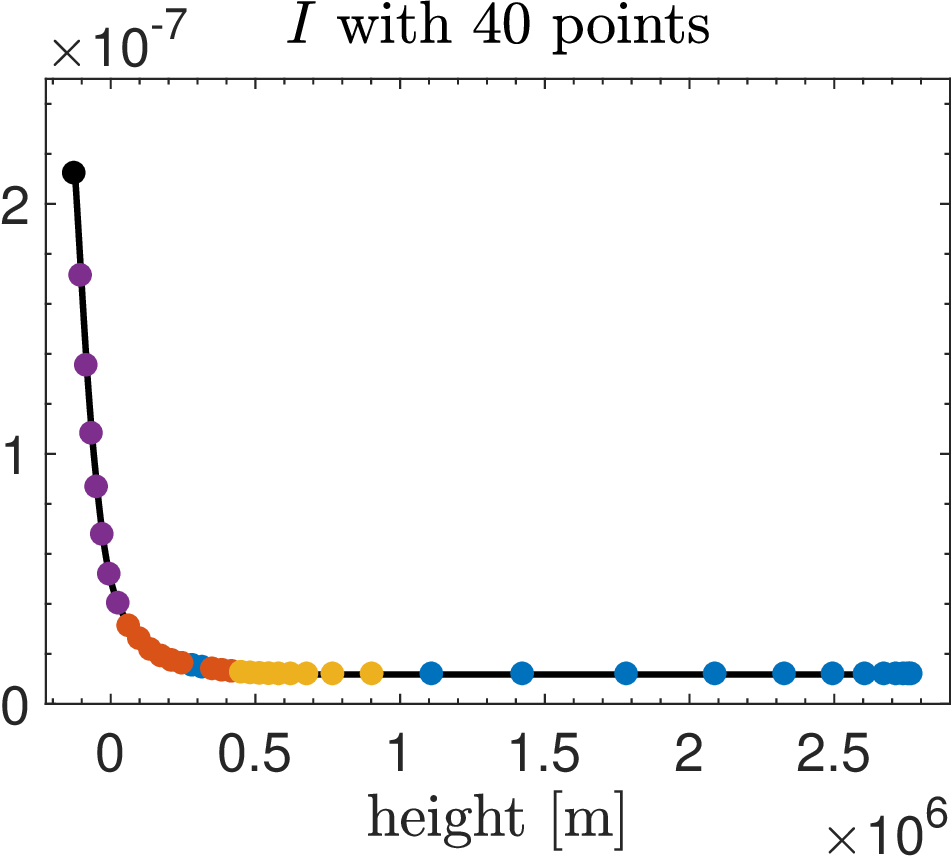}
\includegraphics[width=0.24\textwidth, height=4cm]{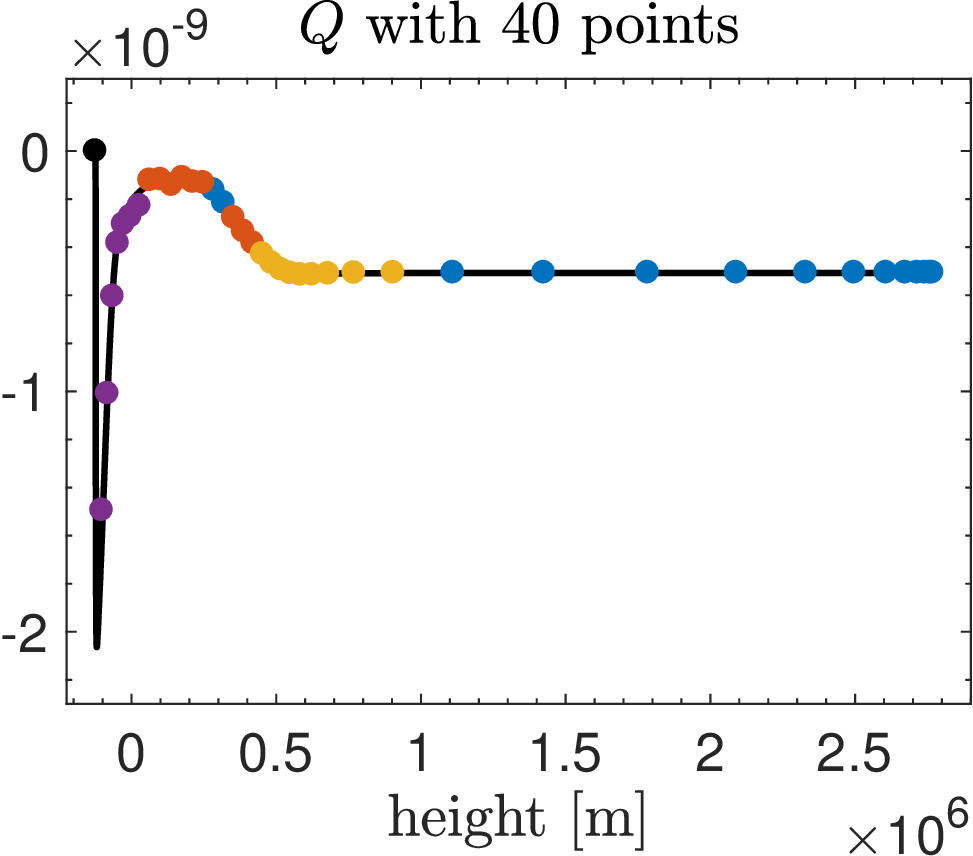}
\includegraphics[width=0.24\textwidth, height=4cm]{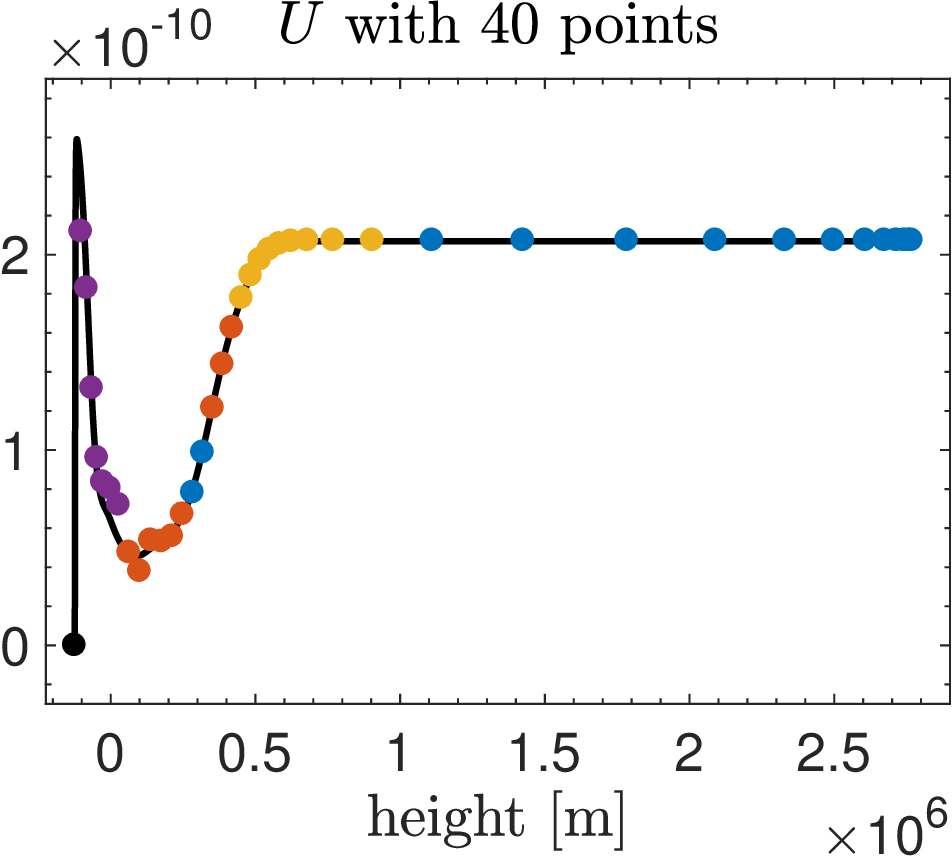}
\includegraphics[width=0.24\textwidth, height=4cm]{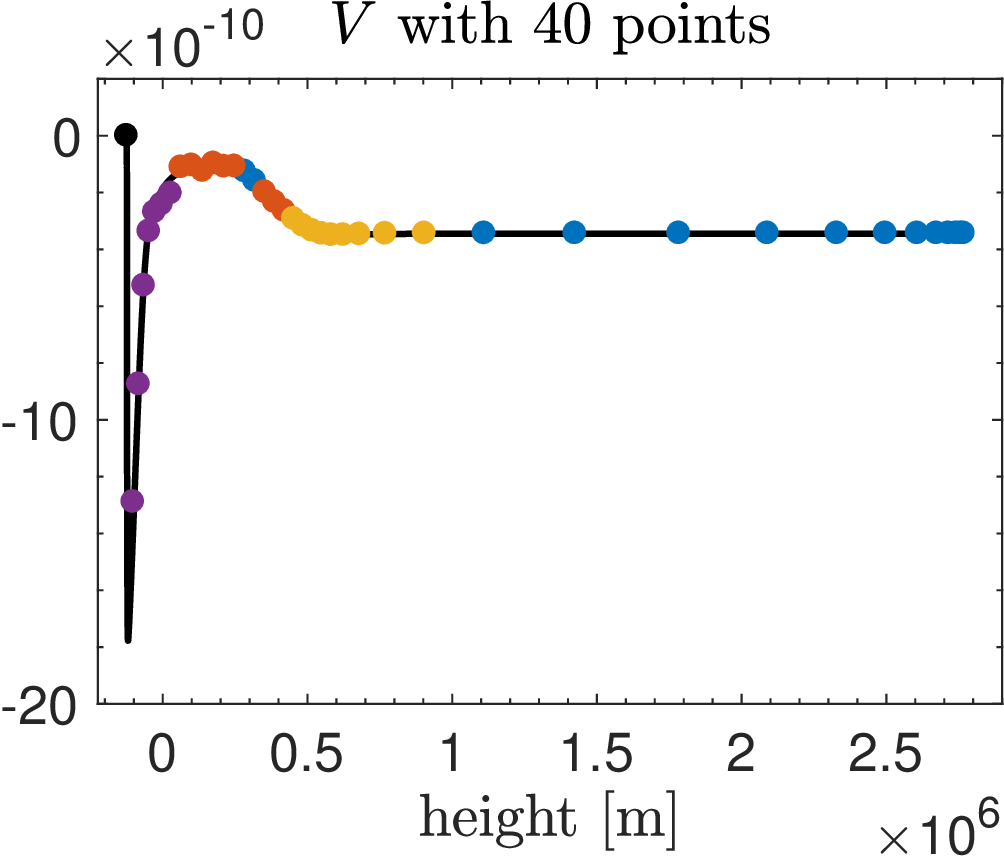}
\\ \vspace{0.3cm}
\includegraphics[width=0.24\textwidth, height=4cm]{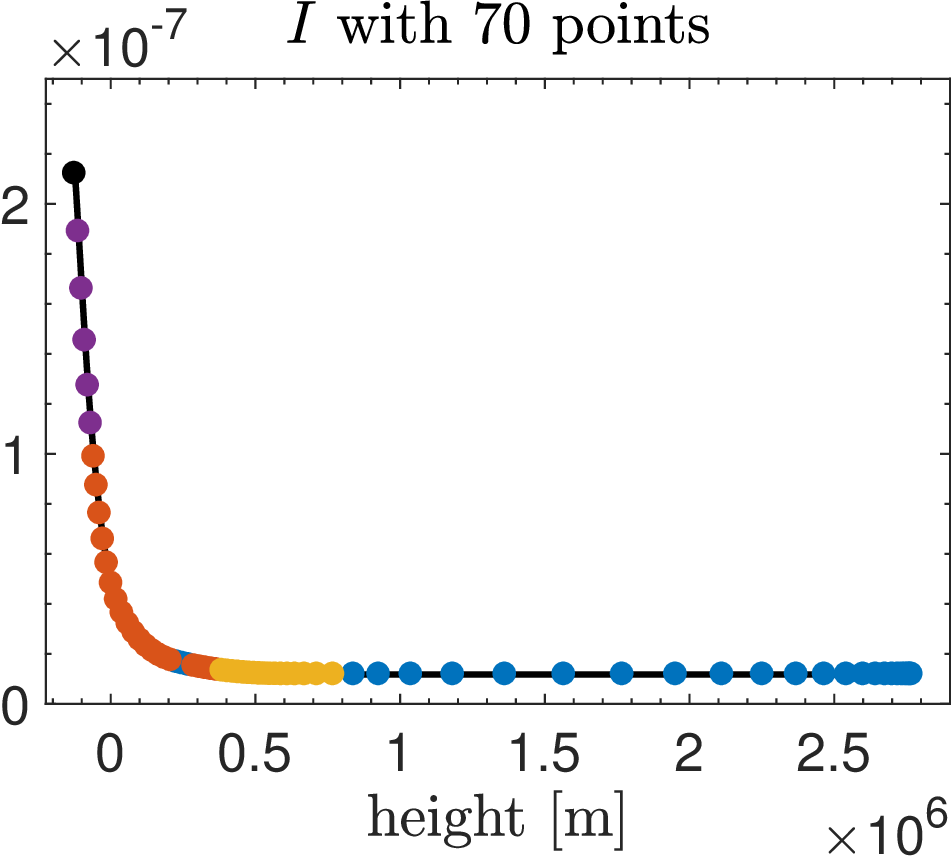}
\includegraphics[width=0.24\textwidth, height=4cm]{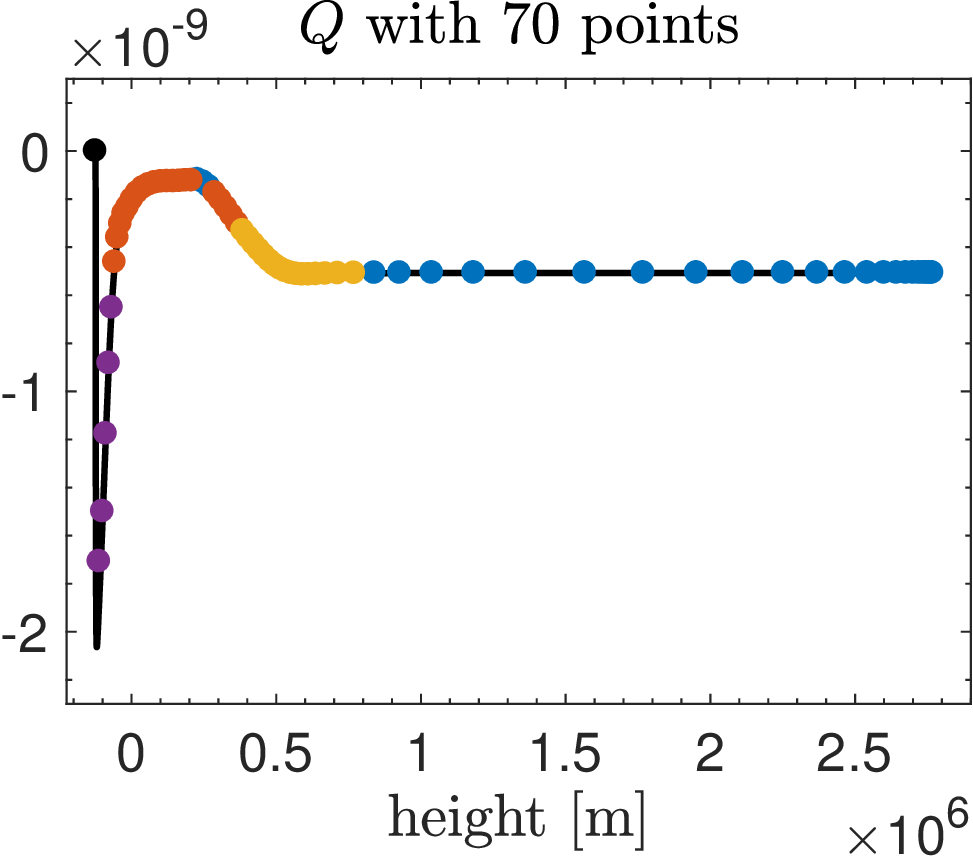}
\includegraphics[width=0.24\textwidth, height=4cm]{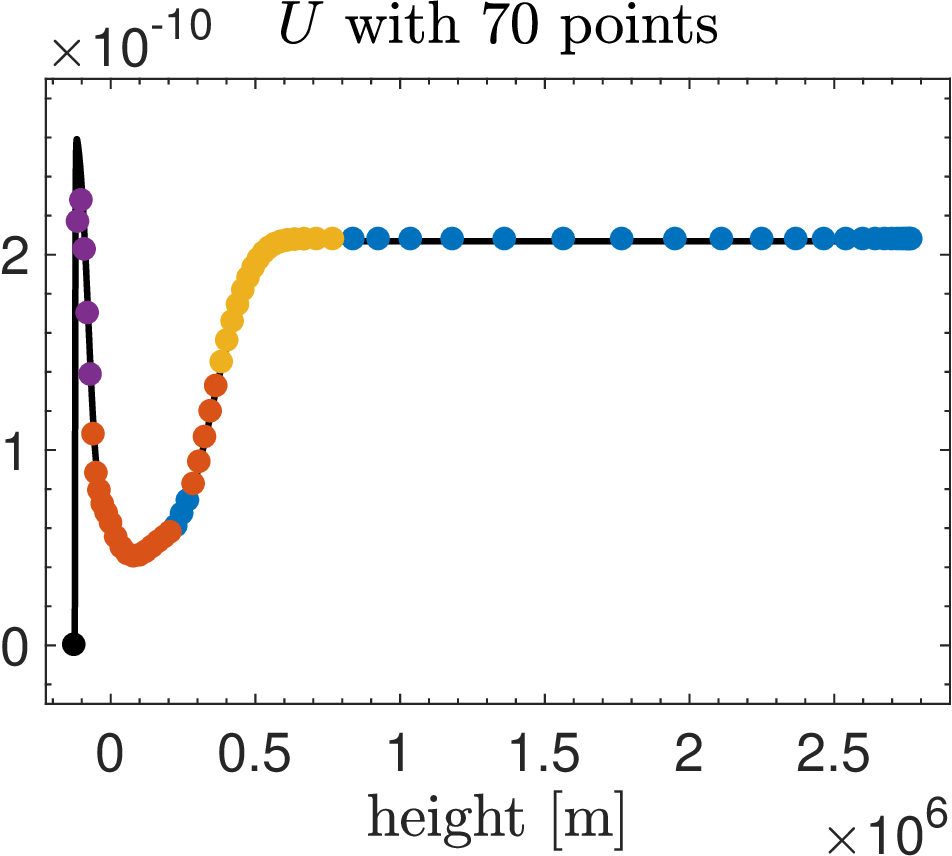}
\includegraphics[width=0.24\textwidth, height=4cm]{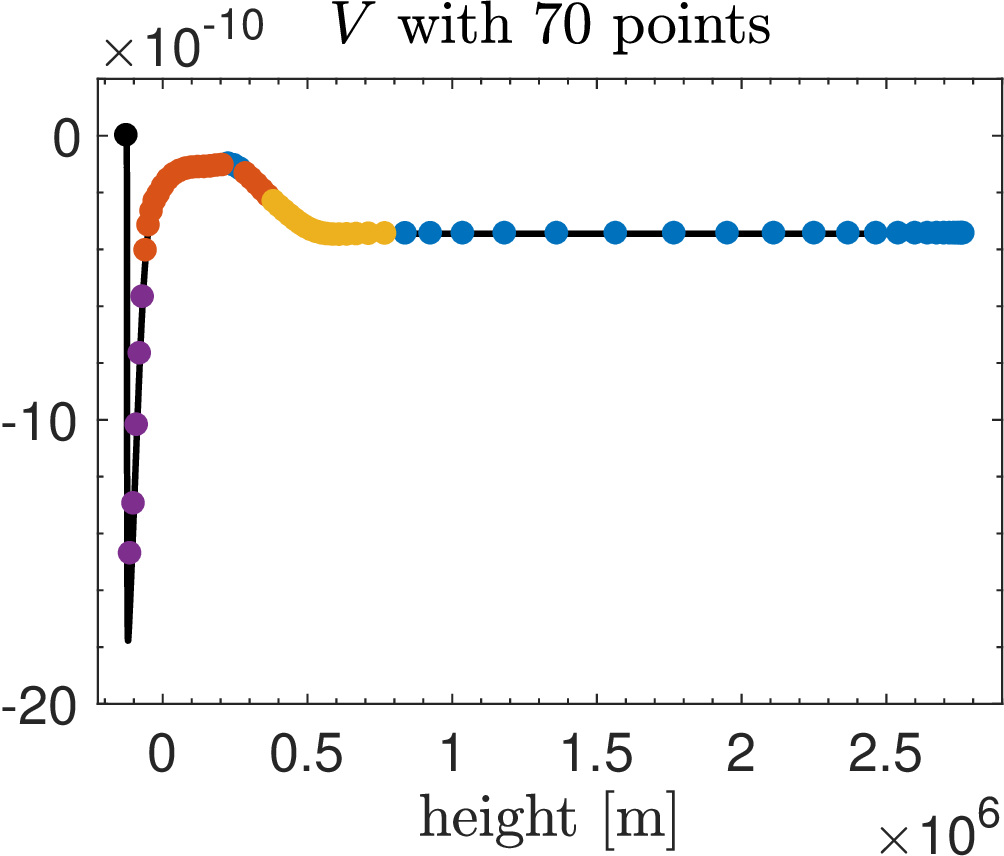}
\\ \vspace{0.3cm}
\includegraphics[width=0.24\textwidth, height=4cm]{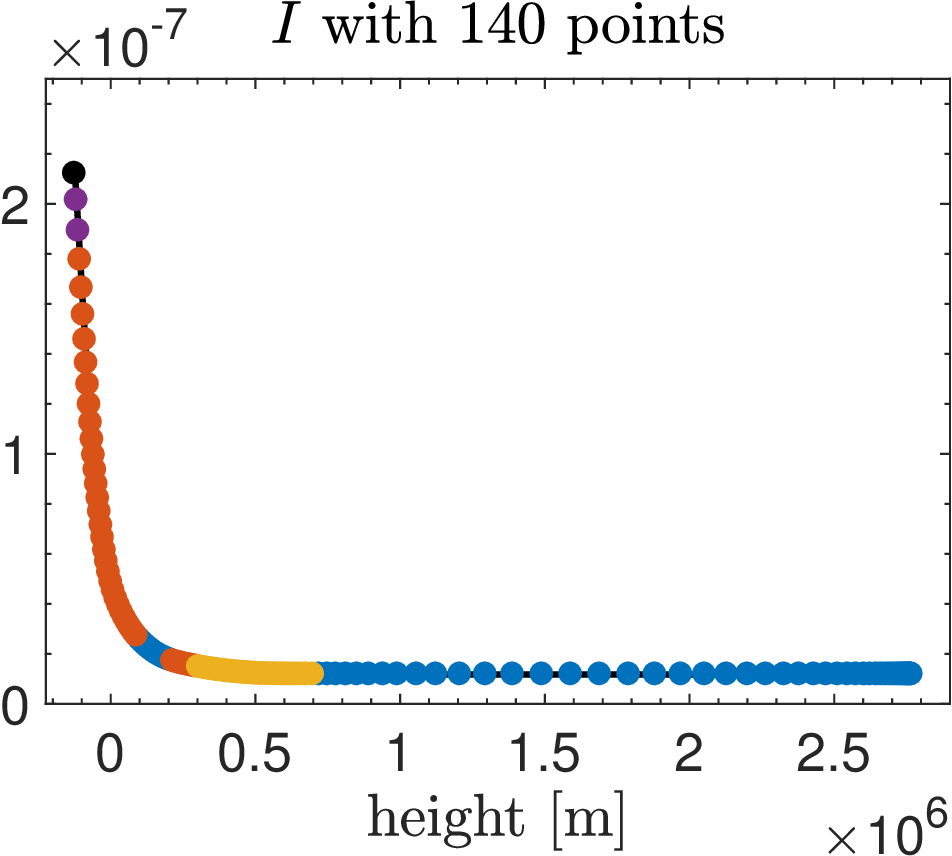}
\includegraphics[width=0.24\textwidth, height=4cm]{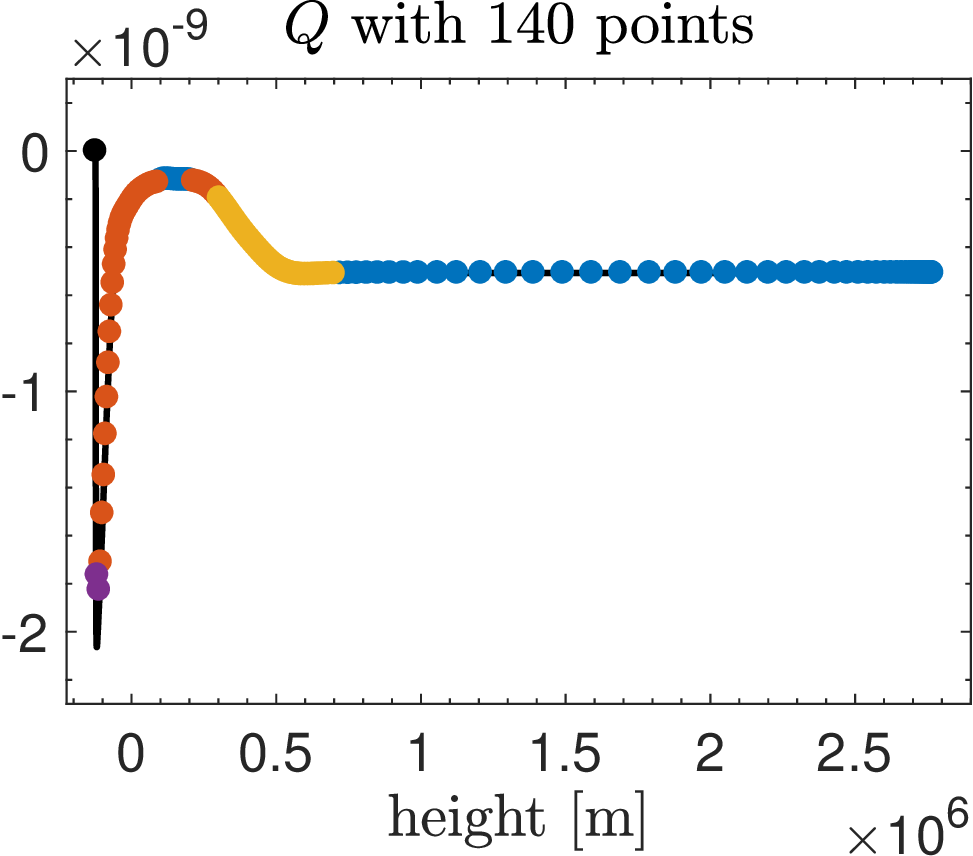}
\includegraphics[width=0.24\textwidth, height=4cm]{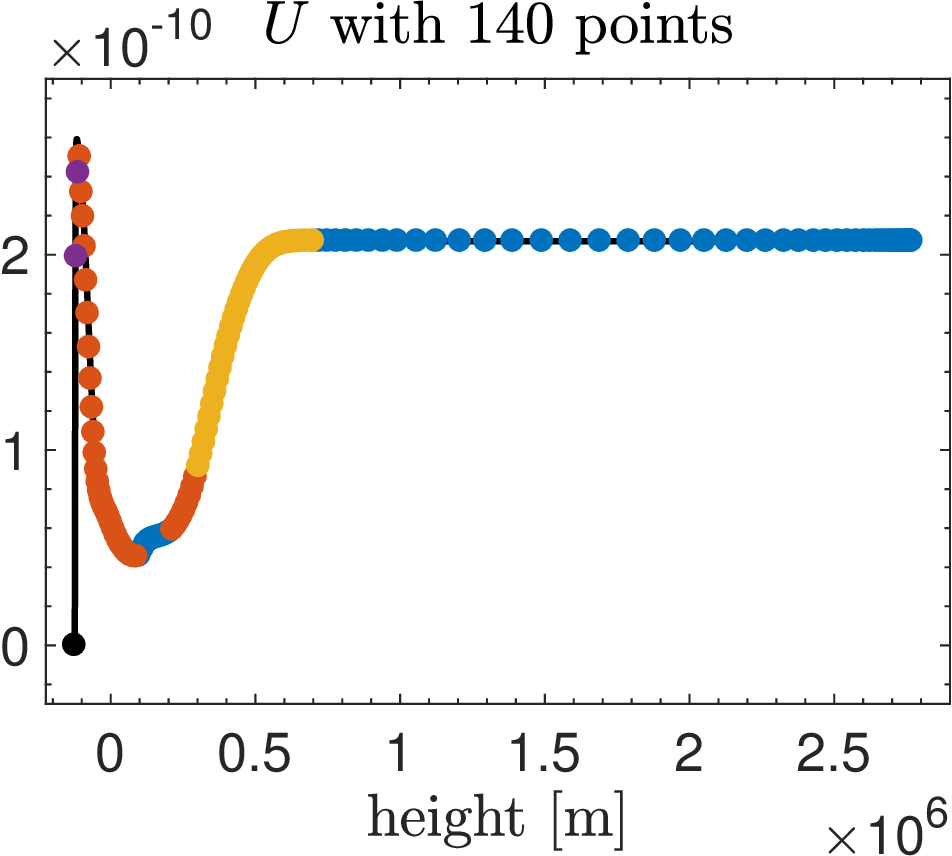}
\includegraphics[width=0.24\textwidth, height=4cm]{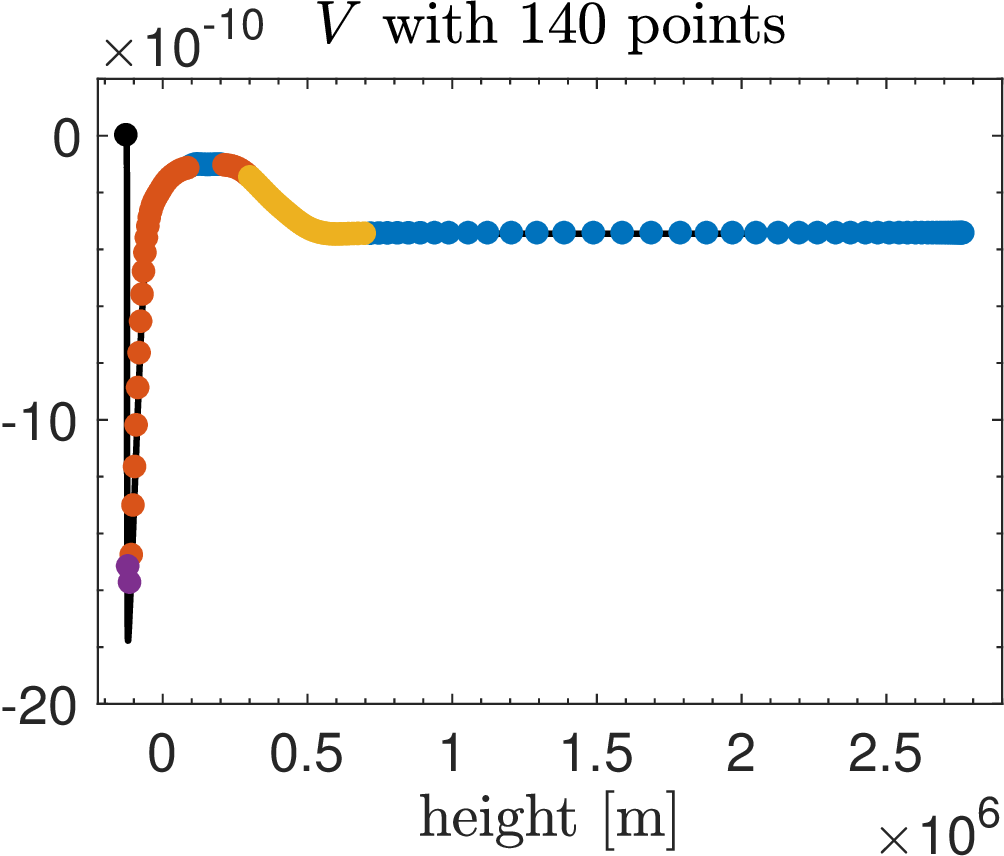}
\\ \vspace{0.3cm}
\includegraphics[width=0.24\textwidth, height=4cm]{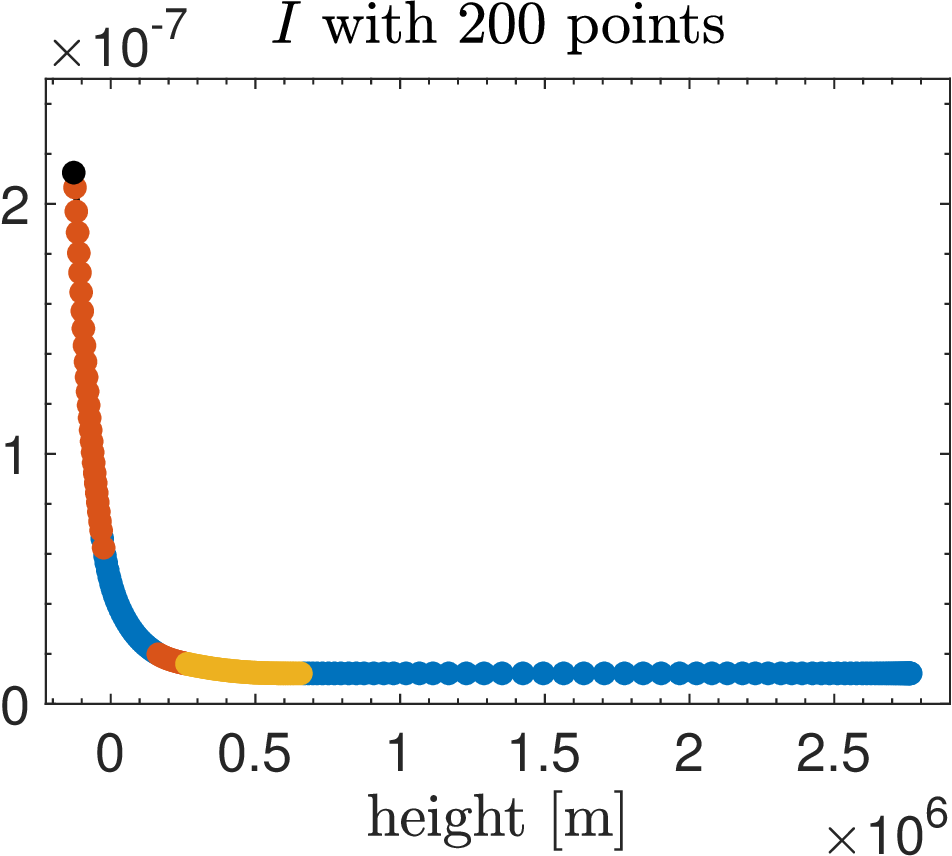}
\includegraphics[width=0.24\textwidth, height=4cm]{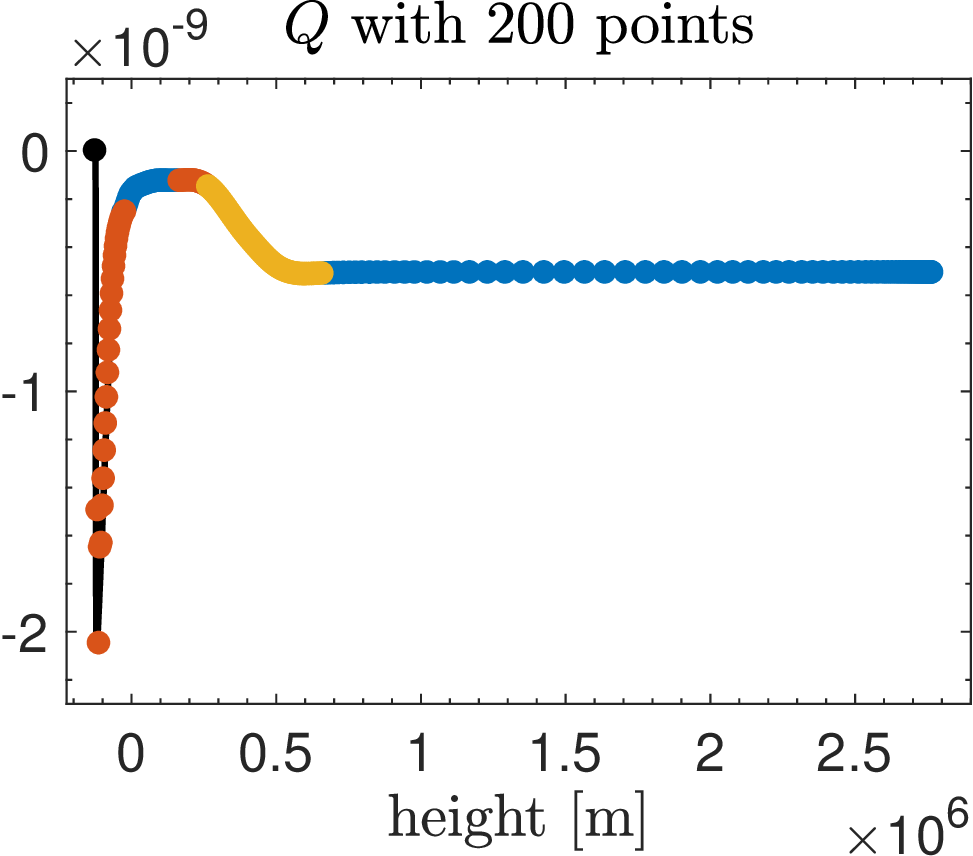}
\includegraphics[width=0.24\textwidth, height=4cm]{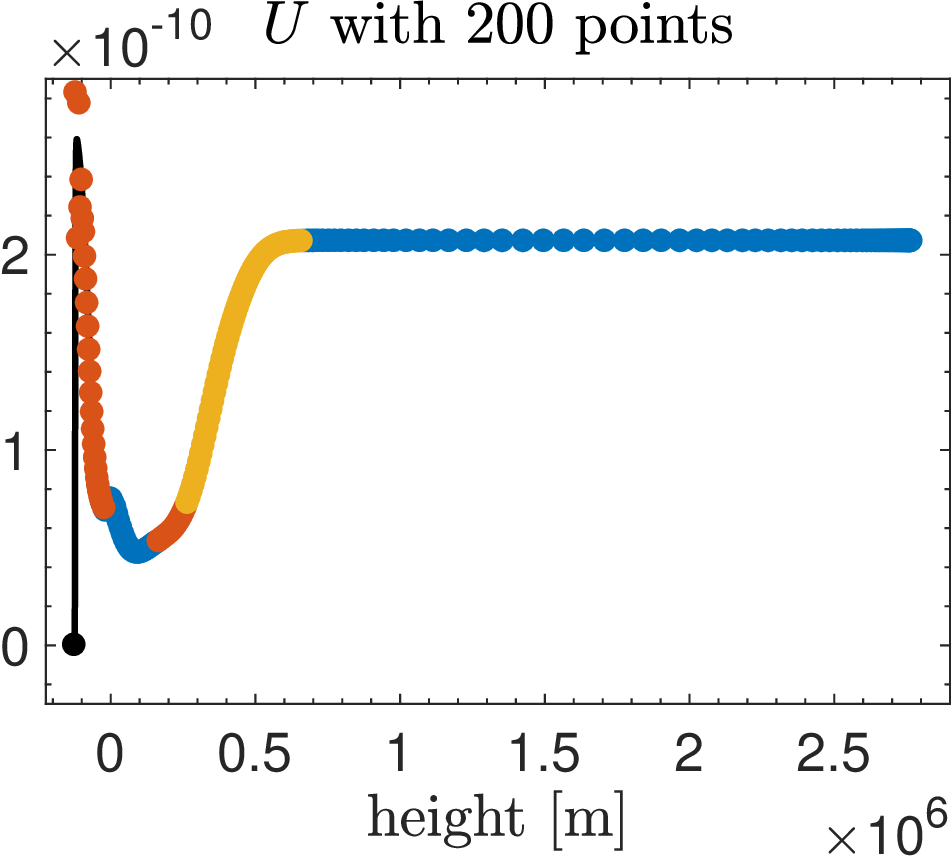}
\includegraphics[width=0.24\textwidth, height=4cm]{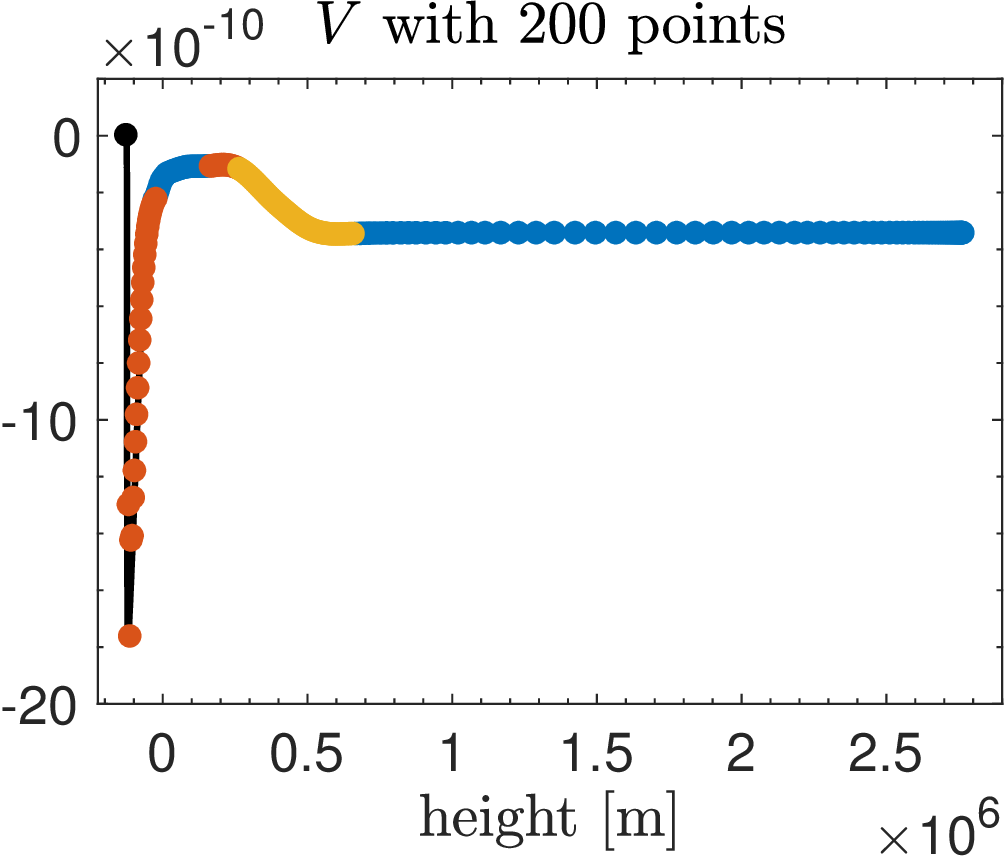}
\caption{
Each row displays the evolution of the Stokes {components}
along the vertical {direction} for the  
Fe~{\sc i} line at 6301.50 {\rm \AA} in the proximity of the line core frequency.
{The Stokes profiles have been computed using a FALC model atmosphere
on a sequence of increasingly refined grids.}
The approximated solution is calculated using the pragmatic approach described in Section~\ref{sec:bestsolver}.
The black line depicts the reference solution, which has been computed with $\Psil$ on a very fine grid.
The initial condition is $\mathbf{I}Ê= (2.121\cdot 10^{-7}, 0, 0, 0)^\top$.
Different dot colors correspond to integrations with a different method.
Blue dots indicate the use of $\Psie$, yellow dots of $\Psia$,  
orange dots of $\Psia$ with conversion to optical depth, and purple dots of $\Psil$.  
The algorithm switches between the different methods depending on
the stability criteria. The use of $\Psie$ increases with the refinement of the grid.}
\label{fig:pragmatic}
\end{figure*}

A practical example is given by Figure~\ref{fig:pragmatic}, which shows the evolution of the approximate Stokes vector
{for the Fe~{\sc i} line at 6301.50 {\rm \AA}}
computed with a FALC atmospheric model \citep{fontenla1993}
{supplemented} with a constant magnetic field\footnote{The values of $\mathbf{K}$  
and $\boldsymbol\epsilon$ have been computed with the RH code of \citet{uitenbroek2001}.}.
The different rows refer to computational  
grids of increasing refinements and the approximate solution is calculated by
the pragmatic numerical scheme suggested above.

The method $\Psil$ is used if there is an eigenvalue
whose real part is $<-7/\Delta s$ (at $s_i$ or $s_{i+1}$).
The method $\Psie$ is used if $\phi_{\Psie}<0.6$
or if the real part of both eigenvalues (at $s_i$ and $s_{i+1}$)
is $>-10^{-3}$.
The method $\Psia$ is converted to optical depth if $\vert\phi_{\Psia}\vert>0.8$.
These parameters should not be considered as an ultimate choice,
but they provide a concrete example.  
However, repeating the experiments with
similar choices of parameters delivers similar results.
The reference solution is computed using the implicit Euler method on
a grid that contains 9999 points.

The experiments show that the pragmatic strategy effectively switches among the methods,
delivering physically meaningful approximations independently from the coarseness
of the grid. As predicted by the analysis, the use of $\Psil$ (purple dots) decreases
with the refinement of the grid:
{it is replaced 
by $\Psia$ (yellow and orange dots), which is in turn replaced by $\Psie$ (blue dots).}
Table~\ref{tab:percentages} summarizes the use (in percentage) of $\Psie$, $\Psia$ {(without and with conversion to optical depth)},
and $\Psil$ for each grid. These values have been approximated
to the second digit.

\begin{table}[htb!]
\caption{Statistic of the pragmatic strategy.}
\centering
\begin{tabular}{|c|c|c|c|c|}
\hline
\emph{\# of grid points} & $\Psie$ & \footnote{Without conversion to optical depth.}$\Psia$ & \footnote{With conversion to optical depth.}$\Psia$ & $\Psil$\\ 
\hline
40 & 36\% & 23\% & 23\% & 18\%\\
70 & 38\% & 30\% & 25\% & 7\%\\
140 & 43\% & 27\% & 28\% & 2\%\\
200 & 51\% & 20\% & 29\% & 0\%\\
\hline
\end{tabular}
\label{tab:percentages}
\end{table}

Although not shown, one must point out that the use of $\Psil$ is necessary in order to deal
with the stiffness of optically thick cells.
This is partly visible in the fourth row, where
$\mathbf{I}_3$ shows overshoots.  
A similar numerical experiment based on $\Psie$ and
$\Psia$ only presents oscillations in the {spatial} region $[-0.12\cdot 10^{5}, 2.5\cdot 10^{5}]$ 
if the grid is too coarse.

\begin{figure*}
\centering
\includegraphics[width=0.24\textwidth, height=4cm]{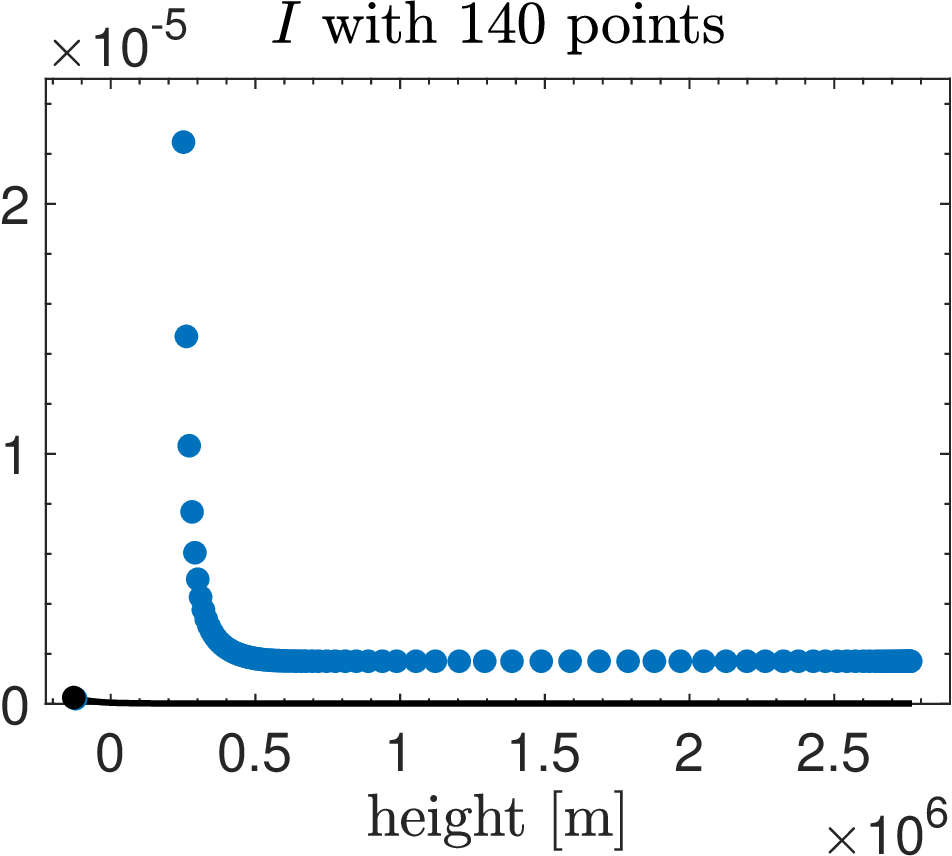}
\includegraphics[width=0.24\textwidth, height=4cm]{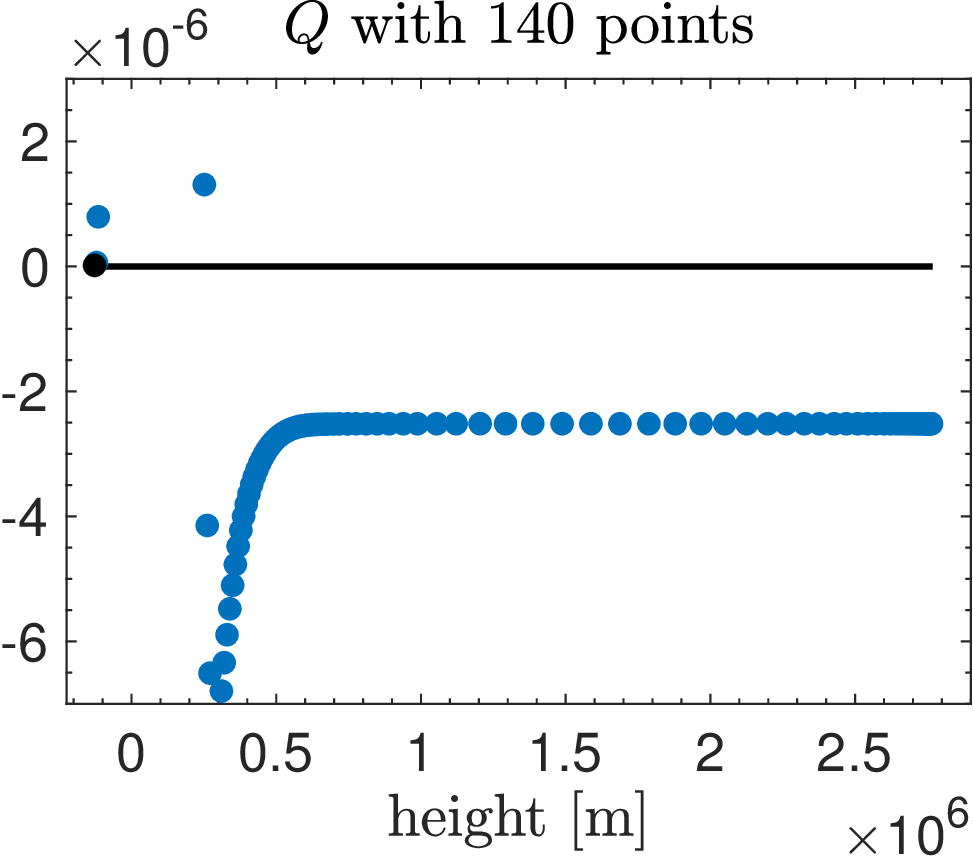}
\includegraphics[width=0.24\textwidth, height=4cm]{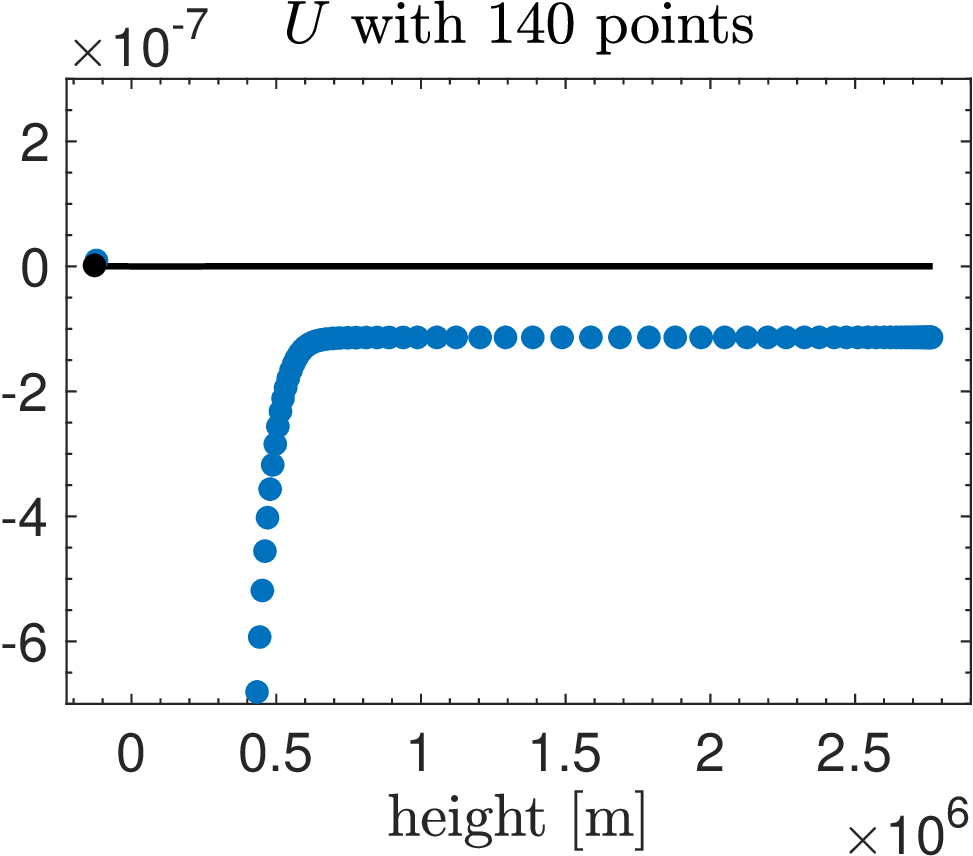}
\includegraphics[width=0.24\textwidth, height=4cm]{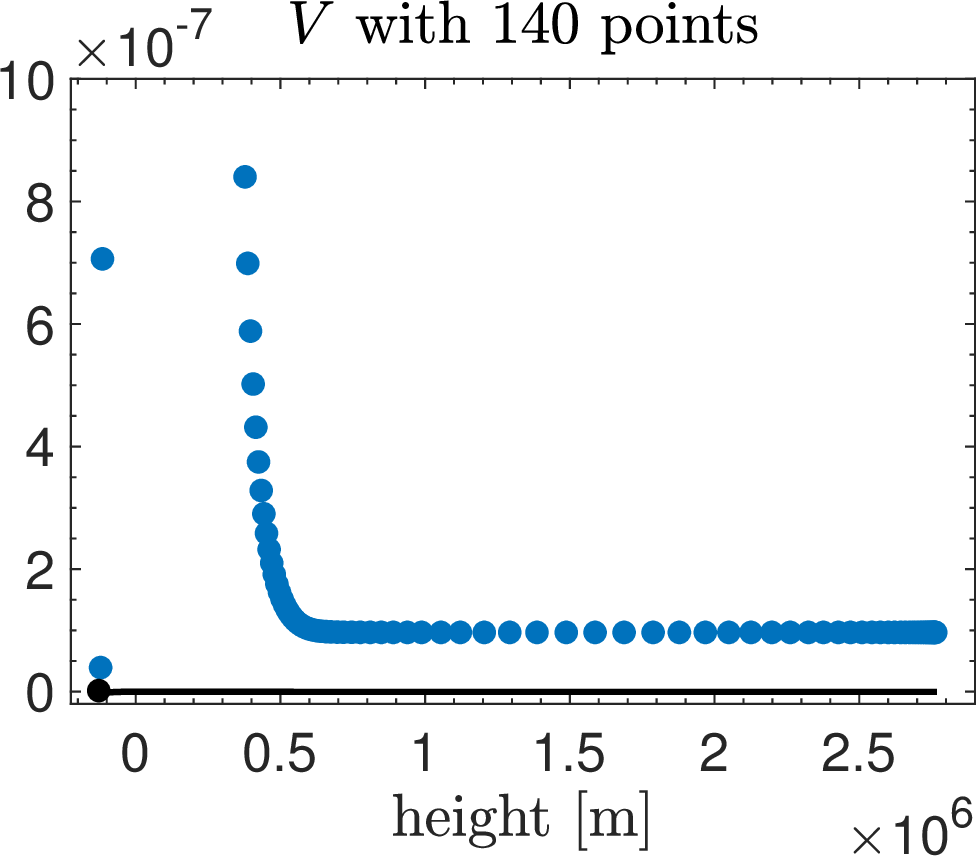}
\\ \vspace{0.3cm}
\includegraphics[width=0.24\textwidth, height=4cm]{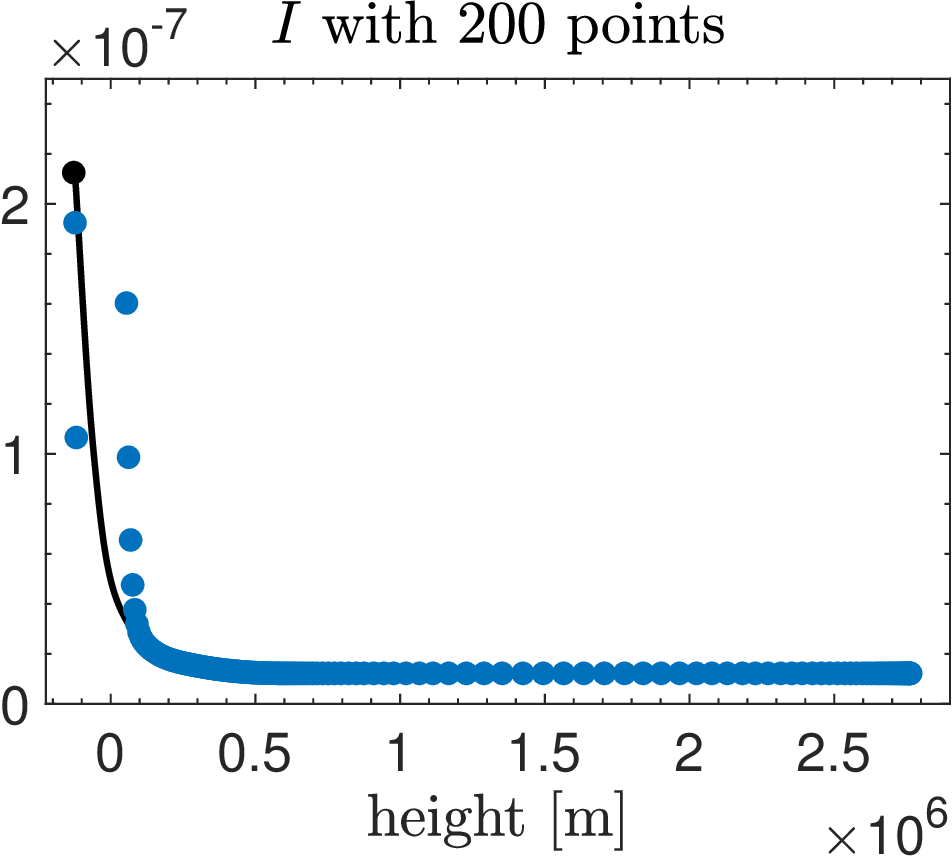}
\includegraphics[width=0.24\textwidth, height=4cm]{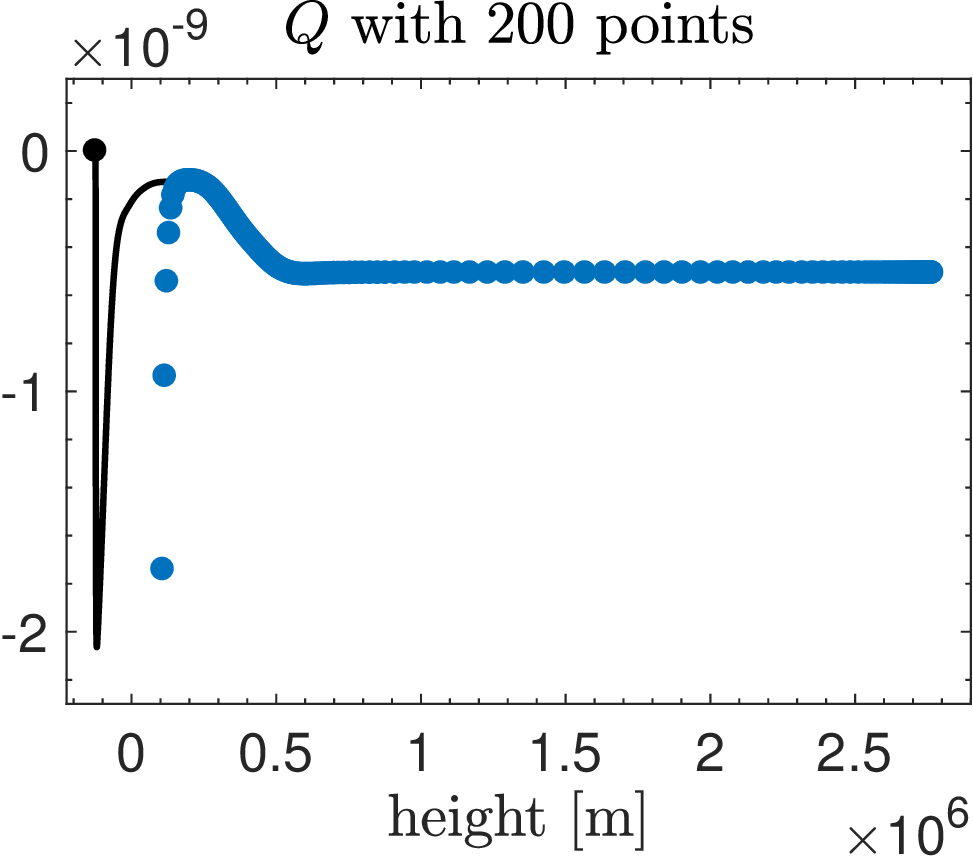}
\includegraphics[width=0.24\textwidth, height=4cm]{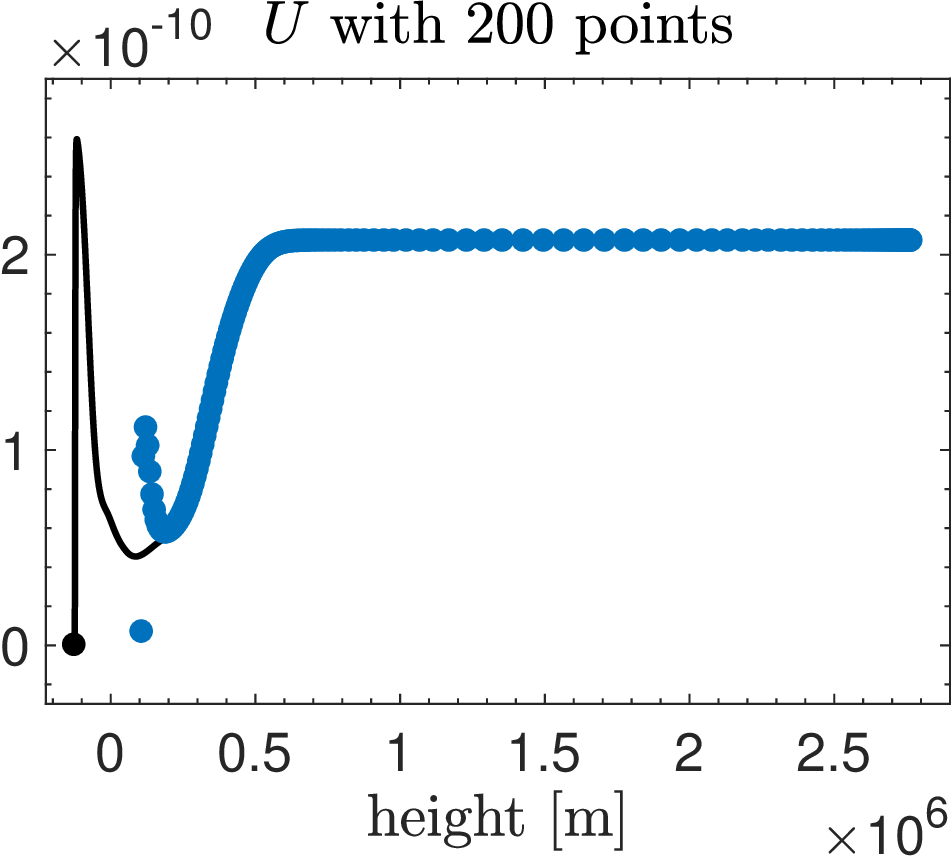}
\includegraphics[width=0.24\textwidth, height=4cm]{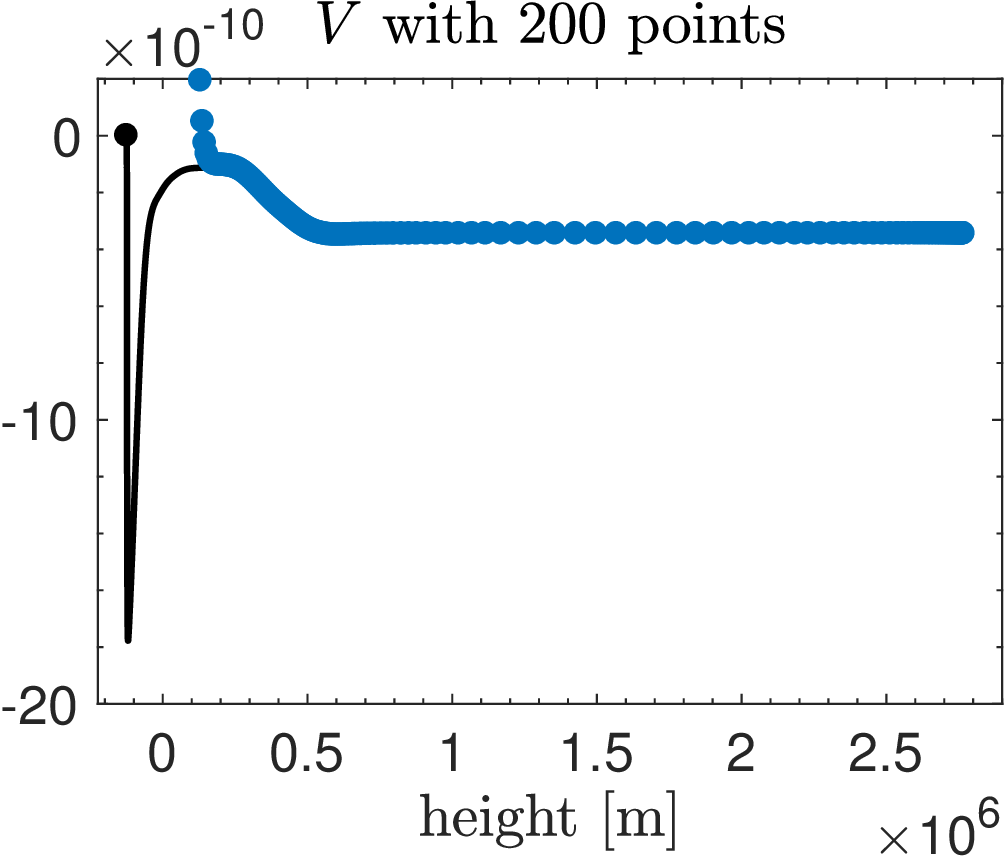}
\caption{
Repetition of the experiment from Figure~\ref{fig:pragmatic}, but using solely
Heun's method to approximate the evolution of the Stokes components.
Calculations
are clearly affected by numerical instabilities.
}
\label{fig:pureHeun}
\end{figure*}

For comparison, Figure~\ref{fig:pureHeun} shows the numerical evolution
of the Stokes vector when this is computed, relying solely on $\Psie$.
With 140 points, this numerical solution is completely spurious because of numerical instability. 
With 200 points, the result is physically correct
only after a certain depth. In particular, in the depth region $[-0.12\cdot 10^{5}, 2.5\cdot 10^{5}]$,
this numerical solution oscillates wildly and the relative error with respect to the reference solution
is of the order of $10^6$.

Finally, using the bound \eqref{eq:spectralradius} on the spectral radius
instead of computing the eigenvalues to decide which method to employ
delivers similar results and is computationally (slightly) cheaper.
\section{Supplemental remarks}\label{sec:sec7}
This section provides two additional considerations concerning the stability of the formal solution of the polarized radiative transfer.
\subsection{Stability of DELO methods}
DELO methods belong to the class of exponential integrators: aiming at removing stiffness from the problem,
the DELO strategy analytically integrates the diagonal elements of the propagation matrix \citep{guderley1972}.
\cite{rees+al1989} first proposed the application of this technique to Equation~\eqref{eq:RTE},
which has been very successful thanks to its stability properties. 
For this reason, the DELO strategy has since been chosen to develop higher-order methods:
e.g., the DELO-parabolic \citep{murphy1990,janett2017a} and the DELO-B\'ezier \citep{delacruz_rodriguez+piskunov2013} methods.
DELO methods are currently widespread for the numerical evaluation of Equation~\eqref{eq:RTE}.

The DELO strategy relies on  
the spatial scale conversion given by Equation~\eqref{eq:conversion_opt_depth}
({which potentially introduces} numerical errors) 
and it deals with the modified propagation matrix $\pmb{\mathscr K}=\mathbf K/\eta_I-\mathbf{1}$,
where $\mathbf 1$ represents the $4\times 4$ identity matrix. 
The stability functions of the DELO-linear and the DELO-parabolic methods satisfy condition~\eqref{eq:Lstable}.
When the norm of the matrix $\pmb{\mathscr K}$ tends to zero {(e.g., for a diagonal matrix $\mathbf K$)},
DELO methods tend to $A$-stability \citep{janett2017a} and, consequently, to $L$-stability.
This fact explains the usual good performance of the DELO-linear method when dealing with very coarse grids
and suggests its suitability as the $L$-stable method $\Psil$ in the pragmatic formal solver described in the previous section.
\subsection{Oscillations in the evolution operator}
Here, a preliminary remark is required. When presenting the fourth-order $A$-stable cubic Hermitian method, \citet{bellot_rubio+al1998}
points to the improper sampling of the oscillations in the evolution operator elements
as a reason for instability and inaccuracy.
In particular, they investigate the case of strong lines, where the cubic Hermitian method flagrantly fails to reliably reproduce the emergent $Q$ and $U$ Stokes components when dealing with coarse spatial grids. 
In light of the stability analysis of Section~\ref{sec:stability}, some considerations can be done.

One starts identifying the origin of the oscillations in the evolution operator elements. The formal solution of Equation~\eqref{eq:RTE} in terms of the evolution operator reads
\begin{equation*}
\mathbf{I}(s) = \mathbf{O}(s,s_0)\,\mathbf{I}(s_0) + 
   \int_{s_{0}}^s \mathbf{O}(s,x)\boldsymbol{\epsilon}(s)\,{\rm d}x\,,
\end{equation*}
where $\mathbf{O}$ is a $4\times 4$ matrix. Under the assumption of a constant propagation matrix $\mathbf K$ in the layer $[s_0,s]$, the evolution operator can be written as
\begin{equation*}
\mathbf O(s,s_0)=e^{-(s-s_0)\mathbf K}\,.
\end{equation*}
If either $\Lambda_+\ne0$ or  $\Lambda_-\ne0$, the evolution operator can be decomposed as
\begin{equation*}
\mathbf O(s,s_0)=\sum_{i=1}^4 e^{-(s-s_0)\lambda^{(i)}}\,\mathbf{N}_i(\boldsymbol{\eta},\boldsymbol{\rho})\,,
\end{equation*}
where $\lambda^{(i)}$ are the eigenvalues given by Equation~\eqref{eigenvalues} and $\mathbf{N}_i$ are known $4\times 4$ matrices \citep[see Appendix~5 of][]{landi_deglinnocenti+landolfi2004}.
From Equation~\eqref{eigenvalues}, one recognizes that
\begin{equation*}
\lambda^{(1)},\lambda^{(2)}\in \mathbb{R}\,,\text{ and }\lambda^{(3)},\lambda^{(4)}\in\mathbb{C}\,.
\end{equation*}
The imaginary part of the eigenvalues $\lambda^{(3)}$ and $\lambda^{(4)}$ induces sinusoidal oscillations in the evolution operator elements, which correspond to the radiative transfer phenomena known as Faraday rotation and Faraday pulsation. These oscillations have spatial frequency $\Lambda_-$, a factor dominated by the anomalous dispersion coefficients (see Table~\ref{tab:lambdas}). In fact, a strong anomalous dispersion vector $\boldsymbol{\rho}$ induces high-frequency oscillations in the evolution operator elements. In particular, the component $\rho_V$ causes a rotation of the direction of maximum linear polarization, whereas the components $\rho_Q$ and $\rho_U$ induce a transformation from linear (circular) to circular (linear) polarization \citep{landi_deglinnocenti+landolfi2004}.

The presence of high-frequency oscillations alone is not a sufficient condition for instability when using an $A$-stable method.
The additional requirement is the variation of the propagation matrix $\mathbf K$ along the integration path.
Moreover, the stability improvement when using denser spatial grids noted by \citet{bellot_rubio+al1998}
is not due to the proper sampling of the oscillations in the evolution operator,
{but} to the {reduction} 
of large variations of the propagation matrix $\mathbf K$ {between} consecutive grid points. 

An illustrative example (not shown here) is given by the numerical evaluation of the formal solution when considering a constant propagation matrix with strong anomalous dispersion coefficients.
In this case, an $A$-stable formal solver (e.g., the trapezoidal method)
integrates Equation~\eqref{eq:RTE} {without} any instability issue. 
 %
\section{Conclusions}\label{sec:sec8}
This paper exposes the stability analysis of the numerical integration of the radiative transfer equation for polarized light.
The main aim is 
to better understand the specific situations where instability issues appear 
and how to deal with them,
rather than prescribing 
the ultimate stability criterion.
This knowledge can be used to devise more robust formal solvers.

The first part focuses on the propagation matrix, identifying different structural properties, such as normality, diagonalizability, spectrum, and spectral radius. 

The second part studies the stability properties of Runge-Kutta methods applied to
Equation~\eqref{eq:RTE}.  Particular attention is paid to the assumptions and the limitations of the stability analysis, emphasizing their relevance in the formal solution for polarized light. 
Special care is paid to better understand the role of spatial variations in the propagation matrix.

It is shown 
that the conversion to the optical depth spatial scale, defined by Equation~\eqref{eq:conversion},
usually {mitigates} variation of the propagation matrix elements along the integration path. 
Appendix~\ref{sec:unpolarizedlight} shows that numerical instabilities due to variations in the eigenvalues are a concrete problem when dealing with polarized light only.
In the scalar case, the conversion to optical depth cancels the variation of the unique eigenvalue along the ray path.
An entire section is dedicated to the numerical conversion to optical depth based on Equation~\eqref{eq:conversion_opt_depth}.
This approximation introduces numerical errors that could lead to a reduced order of accuracy of the formal solver.
In practice, high-order formal solvers require a corresponding high-order numerical evaluation of the integral in Equation~\eqref{eq:conversion_opt_depth}.

Finally, the structure of a paradigmatic pragmatic formal solver is given in terms of a switching technique.
This numerical scheme {chooses} 
between different numerical methods at each step of the integration.
It uses an inexpensive explicit method as long as the integration of the ODE is not limited
by stability requirements and it switches to an implicit method when stiffness appears.
In optically thick cells, the method switches to an $L$-stable method to
correctly replicate exponential attenuations and to avoid numerical oscillations.
The criterion for the switching 
is based either on the eigenvalues or on the spectral radius of the propagation matrix $\mathbf K$. 
The numerical tests are promising: the pragmatic strategy effectively switches among the methods and it
delivers physically meaningful approximations independently from the coarseness
of the grid.

It is important to point out that the stability results presented in this work rely on
assuming that, prior to discretization, the propagation matrix $\mathbf{K}$
and the emission vector $\mathbf{\boldsymbol{\epsilon}}$ are continuous functions.
The effective performance of the pragmatic method on discontinuous atmospheric models
remains to be explored. However, a switching technique based on
choosing numerical methods depending on the local smoothness of the input data
might be suitable to face discontinuities and high gradients.

 %
 %
\acknowledgments 
The financial support by the Swiss National Science Foundation (SNSF) through grant ID 200021\_159206/1 is gratefully acknowledged.
The work of Alberto Paganini was partly supported by EPSRC grant EP/M011151/1.
Moreover, special thanks are extended to L. Belluzzi and O. Steiner for reading and commenting on a previous version of the paper.
The authors also are grateful to the anonymous referee for providing valuable comments that helped improving the article.
\appendix
\section{(Non-)normality of the propagation matrix}\label{appendix:normal}
A real square matrix $\mathbf A$ is normal if 
\begin{equation}
\mathbf A^{T}\mathbf A=\mathbf A\mathbf A^{T}\,,
\label{normality}
\end{equation}
and this is true, e.g., for symmetric or skew-symmetric matrices. Normal matrices are diagonalizable
and their spectrum is stable with respect
to small perturbations of the matrix components \citep{trefethen1999}.

The propagation matrix given by Equation~\eqref{propagation_matrix} satisfies
{\small
\begin{align*}
&\frac{\mathbf K^{T}\mathbf K-\mathbf K\mathbf K^{T}}{2} =\\
&=\begin{pmatrix}
0  &  \eta_U\rho_V\!-\!\eta_V\rho_U &  \eta_Q\rho_V\!-\!\eta_V\rho_Q & \eta_Q\rho_U\!-\!\eta_U\rho_Q  \\
\eta_U\rho_V\!-\!\eta_V\rho_U &  0 &  0 & 0 \\
\eta_Q\rho_V\!-\!\eta_V\rho_Q &  0 &  0 & 0  \\
\eta_Q\rho_U\!-\!\eta_U\rho_Q &  0 & 0 & 0 
\end{pmatrix},
\end{align*}}\noindent
and hence it does not satisfy the normality condition in general. In particular,
$\mathbf K$ is normal if and only if 
\begin{equation*}
\boldsymbol{\eta}\times \boldsymbol{\rho}= \boldsymbol{0}\,.
\end{equation*}
In the cases of vanishing dichroism effects, i.e., $\eta=0$,
or of vanishing anomalous dispersion effects, i.e., $\rho=0$,
one recognizes that the matrix satisfies the normality condition.
\section{Pseudospectrum}\label{appendix:B}
While the spectrum of a matrix is just the set of its eigenvalues, the pseudospectrum also depends on an additional parameter, a small number $\epsilon > 0$. Therefore, one usually refers to the $\epsilon$-pseudospectrum. The $\epsilon$-pseudospectrum $\Lambda_{\epsilon}$ of a square matrix $\mathbf A$ is defined by \citep{trefethen1999}
\begin{equation*}
\Lambda_{\epsilon}(\mathbf A)=\{z\in\mathbb C:\Vert(z\mathbf1-\mathbf A)^{-1}\Vert\ge\epsilon^{-1}\}\,,
\end{equation*}
i.e., it is the set of $z\in\mathbb C$ that are eigenvalues of some matrix $\mathbf A + \mathbf E$ with $\Vert\mathbf E\Vert \le \epsilon$. Here, the matrix $\mathbf E$ acts as a small perturbation to $\mathbf A$. Therefore, the pseudospectrum $\Lambda_{\epsilon}(\mathbf A)$ always contains the $\epsilon$-neighborhood of the spectrum $\Lambda(\mathbf A)$. If one perturbs a normal matrix by a perturbation of operator norm at most $\epsilon$, then the spectrum moves by at most $\epsilon$. However, for a non-normal matrix, the pseudospectrum may be widely dispersed and a small perturbation may induce the spectrum to {change} a lot. 

In order to better understand the concept of pseudospectra, one usually produces approximate pictures of the boundaries of $\Lambda_{\epsilon}(\mathbf A)$ for various values of $\epsilon$, modifying $\mathbf A$ by small random perturbations and looking at the spectra of these perturbations. The standard algorithm is to evaluate the smallest singular value\footnote{The singular values of a matrix $\mathbf A$ are the absolute values of the eigenvalues of the matrix $\mathbf A^\top\mathbf A$.} of the matrix $\mathbf B=z\mathbf1-\mathbf A+\mathbf E$ for different values of $z$ on a grid in the complex plane and then generate a contour plot from this data \citep{trefethen1999}. Different explicit examples are given in Figure~\ref{pseudospectrum}.

These pictures are particularly useful for the stability analysis. \citet{higham1993} conclude that numerical instability around $t$ in Equation~\eqref{IVP1} occurs when the pseudospectra of the frozen matrix $\Delta t\mathbf A$ fail to fit the stability region of the numerical method.
Therefore, one usually analyzes the stability of a numerical method by overplotting its stability region and the boundaries of $\Lambda_{\epsilon}(\Delta t\mathbf A)$ for different values of $\epsilon$, as presented in Figure~\ref{pseudospectrum}.
Note that the stability condition based on pseudospectra is more conservative, or less liberal, with respect to the one based on eigenvalues.

\section{Stability for scalar formal solutions}\label{sec:unpolarizedlight}
The transfer of unpolarized light is described by the first-order inhomogeneous scalar ODE  
\citep{mihalas1978}
\begin{equation}
  \frac{\rm d}{{\rm d} s} I(s) = -\eta_I(s) I(s) + \epsilon(s)\,,
\label{eq:scalar_RTE}
\end{equation}
where $I$ is the specific intensity, $\eta_I$ is the absorption coefficient,
and $\epsilon$ is the emissivity.
To simplify the notation, the frequency dependence of these
quantities is omitted.

In terms of the optical depth $\tau$, Equation~\eqref{eq:scalar_RTE} reads
{\small
\begin{equation}
  \frac{\rm d}{{\rm d} \tau} Z(\tau) = -Z(\tau)  + \frac{\epsilon(g^{-1}(\tau))}{\eta_I(g^{-1}(\tau))}= -Z(\tau)  + S(g^{-1}(\tau))\,,
  \label{eq:scalar_RTE_od}
\end{equation}}\noindent
where $S=\epsilon/\eta_I$ is the so-called source function, $g$ is defined in Equation~\eqref{eq:conversion}, and $Z(\tau) = I(g^{-1}(\tau))$.

The fundamental difference between Equations~\eqref{eq:scalar_RTE} and~\eqref{eq:scalar_RTE_od}
is that in the latter the linear coefficient is constant (and equal to $-1$),
whereas the absorption coefficient $\eta_I$ in Equation~\eqref{eq:scalar_RTE}
depends on $s$.
This implies that to devise a stable numerical scheme
for \eqref{eq:scalar_RTE_od} it is sufficient to follow the discussion presented in
Section \ref{subsec:RKdiag}, whereas for \eqref{eq:scalar_RTE}
one needs to take into account the variations of $\eta_I$ along the ray path, see Section
\ref{subsec:Kvariations}.
In particular, it is sufficient to employ an $A$-stable Runge-Kutta
method to compute stable numerical solutions to Equation~\eqref{eq:scalar_RTE_od},
whereas this may not be sufficient for Equation~\eqref{eq:scalar_RTE}.

For the sake of completeness, the analytic solution to Equation~\eqref{eq:scalar_RTE_od} is given by
\begin{equation*}
Z(\tau) = Z_0e^{-(\tau-\tau_0)} + \int_{\tau_0}^\tau e^{-(x-\tau_0)}
S(g^{-1}(x))\,{\rm d} x\,.
\end{equation*}
Therefore,
\begin{align*}
I(s) &=I_0e^{-(g(s)-g(s_0))} + \int_{s_0}^{s} e^{-(g(y)-g(s_0))}\epsilon(y)\,
{\rm d} y\,,
\end{align*}
which can be approximated in a stable manner by replacing $g$ and the
integral on the right-hand side with numerical approximations.
In this case, monotonicity of $g$ guarantees $L$-stability.

 %
\bibliographystyle{apj} 
\bibliography{bibfile3}
\end{document}